\documentclass[11pt]{article}
\pdfoutput = 1
\usepackage[utf8]{inputenc}

\usepackage[top=1in,bottom=1in,left=1in,right=1in]{geometry}
\makeatletter \g@addto@macro\@floatboxreset\centering \makeatother
\usepackage{color}
\definecolor{darkgreen}{rgb}{0,0.5,0}
\definecolor{darkblue}{rgb}{0,0,0.6}
\definecolor{purple}{rgb}{0.4,.2,0.7}
\usepackage[noEucal]{main}
\usepackage{aas_macros,empheq,fancybox,graphicx,multirow}
\usepackage{aas_macros}
\usepackage{dsfont}
\usepackage{esint}
\usepackage{float}
\usepackage{appendix}
\usepackage{mathabx}
\usepackage{hhline}
\usepackage{tensor} 
\usepackage{footmisc}
\usepackage{caption}
\usepackage{relsize}
\begin{document}

\thispagestyle{empty}

\begin{center}
    ~
    \vskip10mm

     {\LARGE  {\textsc{Lorentz Symmetry and IR Structure \\
     \vspace{10pt}
     of The BFSS Matrix Model}}}
    \vskip10mm
    
Adam Tropper and  Tianli Wang \\
    \vskip1em
    {\it
        Center for the Fundamental Laws of Nature,
Harvard University, Cambridge, Massachusetts 02138, USA\\ \vskip1mm
         \vskip1mm
    }
    \vskip5mm
    \tt{adam$\_$tropper@g.harvard.edu, twang@g.harvard.edu}
\end{center}
\vspace{10mm}

\begin{abstract}
\noindent

The BFSS matrix model relates flat space M-theory to a large $N$ limit of matrix quantum mechanics describing $N$ non-relativistic D0-branes. M-theory, being a theory of gravity in flat space, has a rich infrared structure that includes various soft theorems and an infinite set of conserved charges associated to asymptotic symmetries. In this work, we ask: to what extent is this infrared structure present in BFSS? We find that all the salient features concerning the infrared structure of M-theory carry over naturally to the quantum mechanics dual. Moreover, we demonstrate that the dual statement of the soft graviton theorem in the matrix model implies that D0-brane scattering amplitudes in BFSS enjoy the full $11$d Lorentz symmetry of M-theory, a claim which has been long anticipated. We also offer several first-principle consistency checks for our findings, including a computation of the soft theorem which does not presuppose the BFSS duality and a non-trivial match between several known symmetries of M-theory and BFSS that appear naturally in this formalism. These calculations give non-perturbative evidence in support of the BFSS duality as a model of flat space holography.  

\end{abstract}
\pagebreak

\setcounter{tocdepth}{2}
{\hypersetup{linkcolor=black}
\small
\tableofcontents
}

\newpage

\section{Introduction}

Although the holographic principle has a well-understood manifestation in asymptotically anti-de Sitter (AdS) spacetimes by way of the AdS/CFT correspondence \cite{Weinberg:1965nx}, it has proved challenging to extend this paradigm to asymptotically flat spacetimes. Nevertheless, a holographic description of quantum gravity in asymptotically flat spacetimes seems plausible due to some underlying similarities with AdS. Indeed, the gravitational Hamiltonian is still a boundary term and the entropy of a black hole still scales like its area \cite{Bekenstein:1973ur}.

Historically, there have been two seemingly different approaches to flat space holography. The first is a top-down construction provided by the Banks-Fischler-Shenker-Susskind (BFSS) matrix model, which relates M-theory in flat spacetime to a theory of matrix quantum mechanics describing non-relativistic D0-branes in type IIA string theory \cite{Banks_1997, Susskind:1997cw,Sen:1997we, Polchinski:1999br, Taylor_2001, bigatti1999review, Ydri:2017ncg,Okawa:1998pz,Okawa:1998qk}. This approach endeavors to provide a full non-perturbative duality for M-theory in flat space, and several non-trivial consistency checks have been performed \cite{Becker:1997wh, Becker:1997xw, Kabat:1997sa, Taylor:1998tv,Okawa:1998pz,Okawa:1998qk, Seiberg_1997, deWit:1988wri}. On the other hand, there have only been a few tests of this duality, and most of them concern a very specific setup: $2 \rightarrow 2$ elastic scattering in the eikonal limit. Moreover, recovering the symmetries of M-theory in the dual description has remained a challenging open problem. Even the simplest question of demonstrating that BFSS enjoys 11d Lorentz symmetry has previously been elusive.

The other approach to flat space holography is celestial conformal field theory which aims to use null infinity ($\mathcal{I}^\pm$) as the celestial hologram \cite{deBoer:2003vf,He:2015zea,Pasterski:2016qvg,Strominger:2017zoo,Raclariu:2021zjz,Pasterski_2021}. The kinematics of this approach to flat space holography are extremely well understood, in stark contrast with efforts to analyze BFSS. The Lorentz group manifests itself as the conformal group in the boundary theory, and certain so-called \textit{asymptotic symmetries}, which generalize the Poincar\'e group, play a natural role in the boundary description. Understanding the dynamics of the duality is much more challenging, however. For the most part, this approach to flat space holography is explicitly bottom-up, translating scattering amplitudes in a bulk quantum theory to conformally-covariant correlation functions of operators which live on $\mathcal{I}^\pm$.\footnote{See \cite{Costello:2022jpg, Costello:2022wso, Kar:2022sdc, Kapec_2022, Stieberger:2022zyk, Taylor:2023bzj} for some attempts to give a top-down or non-perturbative description of the duality.}

The overarching goal of this work is to use concepts recently developed in the context of celestial CFT to re-analyze certain  properties of the BFSS matrix model. In particular, we will discuss the infrared structure of M-theory and see how these features are manifested in the matrix theory dual. This approach offers precise predictions about D0-brane quantum mechanics; in so doing, it gives an additional route to test the BFSS duality which has been previously unexplored in the literature. Notably, it gives a universal relation for D0-brane scattering amplitudes and allows one to see how the whole asymptotic symmetry group of M-theory (including, in particular, the full 11d Lorentz symmetry) is manifested in the dual description. Much of the work is focused on translating concrete statements about the infrared structure of M-theory to the D0-brane description by way of the BFSS duality; however, we also offer several first-principle consistency checks of these conjectures. As such, this manuscript gives a new, non-perturbative set of evidence for the BFSS duality as a holographic model of flat space M-theory.

The organization of the article is as follows. In Section \ref{sec: review}, we will review the BFSS duality and survey how supergraviton scattering amplitudes in M-theory are realized in the matrix quantum mechanics. In Section \ref{sec: IR M-theory}, we will discuss the infrared structure of M-theory with a focus on soft theorems, conserved charges, and their corresponding asymptotic symmetries. In Sections \ref{sec: soft theorems}, \ref{sec: conserved charges}, and \ref{sec: asymptotic symmetries}, we will analyze the dual description for these three concepts in the matrix model with an emphasis on demonstrating that BFSS scattering amplitudes enjoy a full 11d Lorentz symmetry as implied by the existence of soft theorems. The fact that the subleading soft graviton theorem implies Lorentz symmetry in generic theories of gravity has only been recently appreciated and is the main conceptual insight that we leverage to argue that $11$d Lorentz symmetry is also enjoyed by the $0+1$ dimensional matrix model \cite{Kapec:2014opa,Kapec:2015vwa}.

This work is a continuation of \cite{Miller:2022fvc}, which gave the original statement of the soft theorem in the matrix model and its relation to asymptotic symmetries of the RR 1-form gauge field in type IIA string theory.

\section{Review of the BFSS Duality}
\label{sec: review}

The BFSS duality suggests that the full non-perturbative description of M-theory in 11d Minkowski spacetime is captured by a certain non-relativistic limit of D0-branes in type IIA string theory \cite{deWit:1988wri,Banks_1997, Susskind:1997cw, Seiberg_1997,Sen:1997we, Polchinski:1999br, Taylor_2001, bigatti1999review, Ydri:2017ncg}. In this section, we review some salient aspects of this duality; familiar readers may safely skip it.

\subsection{The BFSS Matrix Model}

We begin by considering M-theory in flat 11d spacetime in lightcone coordinates $(x^+,x^-,x^I)$ where $x^\pm = \tfrac{1}{\sqrt{2}}(x^0 \pm x^{10})$ are the so-called \textit{longitudinal coordinates} and $x^I$ are the \textit{transverse coordinates} with $I \in \{1,...,9\}.$ 
In this setup, the longitudinal coordinate $x^+$ plays the role of time, and the momentum operator $P^-$ which translates between surfaces of constant $x^+$ is the Hamiltonian.

\subsubsection*{The BFSS Hamiltonian}

The BFSS duality concerns the \textit{discrete lightcone quantization} (DLCQ) of M-theory where the other longitudinal coordinate is compactified and the conjugate momentum is, therefore, quantized 
\begin{equation}
    x^- \sim x^- + 2\pi R \hspace{30pt} p^+ = N/R.
\end{equation}
Here, $N \in \mathbb{N}$ is a positive integer parametrizing the total momentum of a particular M-theory setup. The BFSS duality states that the sector of DLCQ M-theory with total momentum $p^+ = N/R$ is dual to a model of matrix quantum mechanics where the fundamental degrees of freedom are $N \times N$ Hermitian matrices. The dynamics of these matrices are governed by the Hamiltonian 
\begin{equation}
    H_{\text{BFSS}} = \frac{R}{2}\hspace{2pt}\text{Tr}\Bigg[P^I P^I - \frac{1}{2}[X^I,X^J][X^I,X^J] - \Psi^T \Gamma^I[X^I,\Psi]\bigg],
    \label{eqn: BFSS Hamiltonian}
\end{equation}
where $X^I$ is a 9-component vector of $SO(9)$ with $P^I$ its conjugate momentum, $\Psi^\alpha$ is a 16-component spinor of $\text{Spin}(9)$, and $\Gamma^I_{\alpha \beta}$ are the gamma matrices of $\text{Spin}(9)$. All the fields are valued in the adjoint representation of $\frak{u}(N)$. 
The physical Hilbert space of the theory is subject to the Gauss-law constraint which forces physical states to transform as $U(N)$ singlets.

\subsubsection*{BFSS as a Theory of Non-Relativistic D0-Branes}

The BFSS Hamiltonian has a remarkable relationship with D0-brane quantum mechanics \cite{Danielsson:1996uw, Kabat:1996cu, Bachas:1995kx, Sethi_1998, moore2000d, yi1997witten}. The matrices $X^I$ encode the positions of $N$ D0-branes -- an interpretation that becomes especially clear in the classical limit. Classically, the bosonic potential $V \sim -\text{Tr}([X^I,X^J][X^I,X^J])$ is minimized when the $X^I$ matrices are mutually commuting. Therefore, they may be simultaneously diagonalized with eigenvalues tracking the positions of the D0-branes. For example, $N$ non-interacting D0-branes with trajectories $x_1^I(t),...,x_N^I(t)$ are classically described by the diagonal matrices
\begin{equation}
    X^I_{\text{cl.}}(t) = \begin{pmatrix} x^I_1(t) & & \\ & \ddots & \\ & & x^I_N(t)\end{pmatrix}.
\end{equation}
Quantum mechanically, the problem becomes much more subtle; the state is no longer described by an individual matrix, $X^I_{\text{cl.}}$, but rather by a wavefunction over matrices $\Psi[X^I]$. This wavefunction is now supported on matrices of the form
\begin{equation}
    X^I(t) = X^I_{\text{cl.}}(t) + \sqrt{\hbar} \hspace{1pt} \Delta X^I(t),
    \label{eqn: asymptotic states}
\end{equation}
where the $(i,j)$ component of the matrix of quantum fluctuations $\Delta X^I(t)$ is due to interactions between the $i^{th}$ and $j^{th}$ D0-branes via open strings stretched between the two. We have explicitly inserted a factor of $\sqrt{\hbar}$ into the above to emphasize that the fluctuations are quantum corrections. 

\subsection{Asymptotic States and Scattering Amplitudes}
\label{scatteringduality}

Now that both sides of the duality have been stated explicitly, it is important to understand precisely what observables agree between DLCQ M-theory and D0-brane quantum mechanics. Principally, we will be concerned with how S-matrix elements in M-theory are reproduced by corresponding S-matrix elements in BFSS.

In this work, we will only study scattering amplitudes among the supergraviton multiplet in M-theory and the corresponding states in the BFSS matrix model. This is because the infrared structure of M-theory is determined entirely by its massless modes, whose spectrum is captured by 11-dimensional $\mathcal{N} = 1$ supergravity. 

\subsubsection*{Supergraviton Scattering in M-theory}

Supergravitons in M-theory are labelled by their energy-momentum vector, $p^\mu,$ and polarization, $\Theta.$ It is convenient to parametrize the energy-momentum vectors in Cartesian coordinates by an overall scale $\omega \in \mathbb{R}_{\geq 0}$ and a nine-dimensional vector $v^I \in \mathbb{R}^9$ according to
\begin{equation}
    p^\mu = \frac{\omega}{\sqrt{2}}(1+v^{\hspace{1pt}2}, 2v^I, 1-v^{\hspace{1pt}2}),
    \label{eqn: graviton parametrization}
\end{equation}
which satisfies the mass shell condition $p^2 = 0$ for arbitrary $(\omega,v^I).$ The polarization $\Theta$ labels how the state transforms under its little group. The supergraviton multiplet includes a graviton, a three-form gauge field, and a gravitino, which together transform in the $\textbf{44} \oplus \textbf{84} \oplus \textbf{128}$ representation of $SO(9).$

Defining S-matrix elements among supergraviton states in M-theory proceeds in the usual way. One considers the overlap of an in-state composed of $m$ widely separated supergravitons $|p^\mu_1,\Theta_1,...,p^\mu_m,\Theta_m\rangle_{\text{in}}$ and a similarly constructed out state $|p_{m+1}^\mu,\Theta_{m+1},...,p_n^\mu,\Theta_n\rangle_{\text{out}}$
\begin{equation}
    \mathcal{A}_{\text{M}}(p_1^\mu,\Theta_1,...,p_n^\mu,\Theta_n) = ~_\text{out}\langle p_{m+1}^\mu,\Theta_{m+1},...,p_n^\mu,\Theta_n|p_1^\mu,\Theta_1,...,p_m^\mu,\Theta_m\rangle_{\text{in}}.
    \label{eqn: M theory amplitude as overlap}
\end{equation}

\subsubsection*{D0-Brane Scattering in BFSS}

One can also define scattering amplitudes among D0-branes and their bound states in an analogous manner \cite{plefka1998quantum,Plefka_1998,Plefka:1998in,Plefka:1997hm, Becker:1997wh, Becker:1997xw}. Each D0-brane bound state is labelled completely by the quantum numbers $N_j$ (the number of D0-branes in the bound state), $v_j^I$ (the velocity of the D0-brane bound state in the nine spatial directions), and $\Theta_j$ (a $2^8$-dimensional polarization vector transforming in the $\textbf{44} \oplus \textbf{84} \oplus \textbf{128}$ representation of $SO(9)$).

States with multiple, widely separated D0-brane bound states can also be constructed and are labelled respectively by the quantum numbers of the constituent particles. In this way, one can define incoming states $|N_1,v_1^I,\Theta_1,...,N_m,v_m^I,\Theta_m\rangle_{\text{in}}$ and the corresponding outgoing states $|N_{m+1},v_{m+1},\Theta_{m+1},...,N_n,v_n,\Theta_n\rangle_{\text{out}}$ with $ N = N_1+\cdots + N_m = N_{m+1}+\cdots + N_n$. The outgoing states are related to the incoming ones via time evolution with the BFSS Hamiltonian. Scattering amplitudes in the BFSS matrix model are defined by considering the overlap of such states
\begin{equation}
    \begin{split}
        \mathcal{A}_{\text{BFSS}}(N_1,v_1^I,&\Theta_1,...,N_n,v_n^I,\Theta_n) \\
        &= ~_\text{out}\langle N_{m+1},v_{m+1}^I,\Theta_{m+1},...,N_n,v_n^I,\Theta_n|N_1,v_1^I,\Theta_1,...,N_m,v_m^I,\Theta_m\rangle_{\text{in}}.
    \end{split}
\end{equation}

\subsubsection*{The BFSS Duality Dictionary}

It is no coincidence that the quantum numbers describing supergravitons and those describing D0-brane bound states are so similar. Indeed, it is well-known that supergraviton states in DLCQ M-theory are dual to bound states of D0-branes in the matrix model \cite{Banks_1997, Susskind:1997cw, Seiberg_1997, Becker:2006dvp}. The parameters are related via Table \ref{tab: Kinematics Dictionary}. Moreover, the duality asserts that supergraviton scattering amplitudes agree with D0-brane scattering amplitudes when the parameters are related by this duality dictionary
\begin{equation}
    \mathcal{A}_{\text{M}}(p_1^\mu,\Theta_1,...,p_n^\mu,\Theta_n) = \mathcal{A}_{\text{BFSS}}(N_1,v_1^I,\Theta_1,...,N_n,v_n^I,\Theta_n). 
    \label{eqn: BFSS duality}
\end{equation}

\begin{table}
\centering
\small
\begin{tabular}{||c||} 
\hhline{|t:=:t|}
         \hspace{0pt} \textbf{M-Theory} \hspace{80pt} \textbf{BFSS} \hspace{20pt}  \\
\hhline{|:=:|}
          \rule{0pt}{5ex} \hspace{14pt} $p_{j}^+$ \hspace{35pt} $\Longleftrightarrow$ \hspace{38pt} $\mathlarger{\frac{N_j}{R}}$ \hspace{20pt} \\
          \rule{0pt}{5ex} \hspace{32pt} $p_j^I$ \hspace{38pt} $\Longleftrightarrow$ \hspace{24pt} $\sqrt{2}\mathlarger{\frac{N_j}{R}} v_j^I $ \hspace{30pt} \\
          \rule{0pt}{5ex} \hspace{16pt} $p_{j}^-$ \hspace{36pt} $\Longleftrightarrow$ \hspace{32pt} $\mathlarger{\frac{N_j}{R}} v_j^2$ \hspace{20pt} \\
          \rule{0pt}{4ex} \hspace{10pt} $\Theta_j$ \hspace{37pt} $\Longleftrightarrow$ \hspace{38pt} $\Theta_j$ \hspace{20pt} \\
\hhline{|b:=:b|}
\end{tabular}
\caption{Dictionary between the kinematic data of gravitons in M-theory and D0-brane clumps in the matrix model. The relation concerning $p_{j}^-$ is determined using the mass-shell condition for M-theory gravitons $0 = -2p_j^+p_j^- + p_j^I p_j^I$.}
    \label{tab: Kinematics Dictionary}
\end{table}
There have been several previous checks that scattering amplitudes in the two theories agree (e.g. \cite{Becker:1997wh, Becker:1997xw, Kabat:1997sa, Taylor:1998tv,Okawa:1998pz,Okawa:1998qk}); however, explicit calculations have only been performed in the low-energy, elastic limit of supergraviton scattering in M-theory, and many puzzles remain \cite{Helling:1999js,Becker:1997cp,Keski-Vakkuri:1997iht,Taylor_2001}. 

In this article, we assume that BFSS provides an exact duality between scattering amplitudes in M-theory and those in D0-brane quantum mechanics. Specifically, even though a full understanding of BFSS is lacking, we will use Equation \eqref{eqn: BFSS duality} to infer conjectures about the IR structure of the BFSS matrix model. Fortunately, we also provide significant first-principle evidence for this IR structure which is agnostic to the BFSS duality and does not presuppose Equation \eqref{eqn: BFSS duality}. This, in turn, supplies additional consistency checks of Equation \eqref{eqn: BFSS duality} and supports the veracity of the BFSS duality as a whole.

\subsubsection*{The Decompactified M-Theory Limit}

Up to this point, we have been equating amplitudes in the discrete lightcone quantization of M-theory with amplitudes in the BFSS matrix model at finite $N$. To recover scattering amplitudes in uncompactified M-theory, one sends the radius of compactification to infinity while holding the external momentum fixed \cite{Banks_1997, Susskind:1997cw}
\begin{equation}
    R \rightarrow \infty \hspace{30pt} p_j^\mu = \text{const.}
\end{equation}
Using the duality dictionary, we find that this is equivalent to the following scaling limit of the matrix model where the number of D0-branes in a particular bound state becomes infinitely large 
\begin{equation}
    R \sim N \sim N_j \rightarrow \infty \hspace{30pt} v_j^I = \text{const.} \hspace{52pt}
\end{equation}

\section{Infrared Structure of M-theory}
\label{sec: IR M-theory}

It is well-known that all gravitational theories in asymptotically flat spacetime exhibit certain universal features. As M-theory includes gravitons, these universal results are applicable and provide interesting non-perturbative information about the infrared structure of M-theory. In this section, we discuss how supergraviton scattering amplitudes obey precise soft theorems. We also review how these soft theorems are equivalent to Ward identities enforcing the conservation of infinitely many charges associated to asymptotic symmetries.\footnote{Readers familiar with the infrared structure of a general gravitational theory may safely skip this section as the focus is merely on translating certain universal features of gravity to the specific case of M-theory in flat 11-dimensional spacetime.}

\subsection{Soft Theorems}

Consider a generic $(n+1)$-particle scattering amplitude which includes a graviton with momentum $q_s^\mu$ and polarization tensor $\varepsilon^{\mu \nu}$ among its external states. We write this amplitude as $\mathcal{A}_{\text{M}}(q_s^\mu,\varepsilon;p_1^\mu,...,p_n^\mu)$ where $p_1^\mu,...,p^\mu_n$ label the future-directed momenta of the other $n$ particles. We will focus on the case where the other particles are also supergraviton states, though the result holds even outside of this setting. 

The soft graviton theorem demonstrates that this amplitude obeys a universal factorization property in the limit  $\omega_s/\omega_j \rightarrow 0$ where $\omega_s$ (respectively $\omega_j$) is the overall scale of $q_s^\mu$ (respectively $p_j^\mu$) as defined in Equation \eqref{eqn: graviton parametrization} \cite{Weinberg:1965nx,Kapec_2017,He_2019_1,He_2019,Kapec_2022,Marotta:2019cip, Cachazo:2014fwa}. This is the so-called \textit{soft limit}, and one may perform a series expansion of the amplitude in this ratio. The \textit{soft graviton theorem} states
\begin{equation}
    \mathcal{A}_{\text{M}}(q_s^\mu,\varepsilon;p_1^\mu,...,p_n^\mu) = \Big[\mathcal{S}^{(-1)} + \mathcal{S}^{(0)} + \cdots \Big]\mathcal{A}_{\text{M}}(p_1^\mu,...,p_n^\mu) \hspace{30pt} \text{where}  \hspace{30pt} \mathcal{S}^{(n)} ~ \propto ~ \Big(\frac{\omega_s}{\omega_j}\Big)^n
    \label{eqn:Mth SGT}
\end{equation}
and so relates an amplitude with a soft graviton to an amplitude without one. Moreover, it has been demonstrated that $\mathcal{S}^{(-1)}$ and $\mathcal{S}^{(0)}$ take the following universal form \cite{Weinberg:1965nx,Cachazo:2014fwa}
\begin{equation}
   \mathcal{S}^{(-1)} = \frac{\kappa}{2} \hspace{2pt} \varepsilon_{\mu \nu} \sum_{j=1}^n \eta_s \eta_j \frac{p_j^\mu p_j^\nu}{q_s \cdot p_j} \hspace{35pt} \mathcal{S}^{(0)} =  i \frac{\kappa}{2} \varepsilon_{\mu \nu} \sum_{j =1}^n \eta_s \eta_j \frac{p^\mu_j J^{\nu\rho}_j q_{s,\rho}}{q_s \cdot p_j},
\end{equation}
 where $\kappa = \sqrt{32\pi G_N}$ is Newtons's constant, and we have defined $\eta_j = -1$ if particle $j$ is incoming and $\eta_j = +1$ if the particle $j$ is outgoing (similarly for $\eta_s$). $\mathcal{S}^{(-1)}$ is called the \textit{leading soft term} while $\mathcal{S}^{(0)}$ is called the \textit{sub-leading soft term}. Note that the leading soft term has a pole $\propto ~ (\omega_s/\omega_j)^{-1}$, so the amplitude diverges in the soft limit. Finally, $J^{\mu \nu}_j$ is the generator of angular momentum of the $j^{th}$ particle and is composed of both orbital and spin operators
\begin{equation}
    J^{\mu \nu}_j = L^{\mu \nu}_j + S^{\mu \nu}_j.
\end{equation}
In general, there will also be sub-subleading terms in the soft expansion which have positive powers of $(\omega_s/\omega_j)$ in the soft limit; however, these terms are theory dependent and won't play a role in our analysis.

\subsection{Conservation Laws}

The leading and subleading soft graviton theorems have been shown to be equivalent to Ward identities which enforce various symmetries in scattering amplitudes. To this end, it is convenient to express the $n$-particle amplitude as the overlap of incoming and outgoing multi-supergraviton states according to Equation \eqref{eqn: M theory amplitude as overlap}. To simplify the notation, we will use the shorthand $|\text{in}\rangle = |p_1^{\mu},\Theta_1,...,p_m^\mu,\Theta_m\rangle_{\text{in}}$ and $|\text{out}\rangle = |p_{m+1}^{\mu},\Theta_{m+1},...,p_n^\mu,\Theta_n\rangle_{\text{out}}$ to stand in for these states.

Next, we define the creation operator $G_{IJ}^{\hspace{1pt}\text{in} \dagger}(\omega_s,v_s)$ for an incoming graviton state with momentum $q_s^\mu$ parameterized by $(\omega_s,v_s)$ and polarization tensor $\varepsilon^{\mu \nu} = (\varepsilon_{IJ})^{\mu \nu}$. Here, $(\varepsilon_{IJ})^{\mu \nu}$ are defined by 
\begin{equation}
    (\varepsilon_{IJ})^{\mu \nu} = \frac{1}{2}(\varepsilon^\mu_I \varepsilon^\nu_J + \varepsilon^\nu_I \varepsilon^\mu_J) - \frac{1}{9}\delta_{IJ} \varepsilon^\mu_K\varepsilon^{K \nu} \hspace{30pt} \text{with} \hspace{30pt} \varepsilon^\mu_J = (v^J_s,\delta^I_J,-v^J_s).
\label{eqn:polbasis}
\end{equation}
They are symmetric and traceless in the $I,J = 1,...,9$ indices and form a basis for the graviton's 44 polarization states. One may similarly define the operator $G_{IJ}^{\hspace{1pt}\text{out} \dagger}(\omega_s,v_s)$ which is responsible for creating an outgoing graviton state. 

It has been shown that the leading and subleading soft theorems are each equivalent to the conservation of infinitely many charges. This correspondence has been demonstrated explicitly by a number of authors, and we will forgo the derivation in this work \cite{Strominger:2017zoo,He_2015,Kapec:2014opa,Campiglia:2014yka,Kapec_2017,Kapec_2018,Kapec_2022}. Rather, we will merely quote the results.

\subsubsection*{Leading Order}

In particular, one may define charges $Q^{\text{in}}[f]$ and $Q^{\text{out}}[f]$ which are associated to the leading soft theorem and respectively act on in and out states. These charges are parametrized by an arbitrary function $f: \mathbb{R}^9 \rightarrow \mathbb{R}$ and are conserved in all physical processes. This conservation implies the following relation among scattering amplitudes
\begin{equation}
    \begin{split}
        \langle \text{out}|\Big(Q^{\text{out}}[f] - Q^{\text{in}}[f]\Big)|\text{in}\rangle = 0.
    \end{split}
\end{equation}
Both the in and the out charge split up into a sum of two terms: a \textit{hard term} and a \textit{soft term} labelled by subscripts $H$ and $S$ respectively
\begin{equation}
    Q^{\text{in}}[f] = Q^{\text{in}}_H[f] + Q^{\text{in}}_S[f].
    \label{eqn: Qf}
\end{equation}
These charges may be defined by their action on the asymptotic states. Specifically
\begin{equation}
    \begin{split}
        Q_H^{\text{in}}[f] |\text{in}\rangle &= \sum_{j \in |\text{in}\rangle} \omega_j f(v_j) |\text{in}\rangle \\
        Q_S^{\text{in}}[f] |\text{in}\rangle &=  - \frac{1}{4 \kappa} \lim_{\omega_s \rightarrow 0} \bigg[\omega_s\int d^9 v_s ~ f(v_s) ~ \partial_{v_s^I} \partial_{v_s^J} \widetilde{G}^{(-1) \text{in} \dagger}_{IJ}(\omega_s,v_s)\bigg] |\text{in}\rangle,
    \end{split}
\label{eqn:mtheoryleadingcharge}
\end{equation}
where $(\omega_j,v_j)$ label the momenta of the incoming supergraviton states according to Equation \eqref{eqn: graviton parametrization}. The out charges are constructed similarly. Here, $\widetilde{G}^{(-1) \text{in} \dagger}_{IJ}(\omega_s,v_s)$ denotes a certain integral transform of the graviton creation operator $G^{\text{in} \dagger}_{IJ}(\omega_s,v_s)$. This integral transform is known as the \textit{shadow transform} and is discussed in greater detail in Appendix \ref{appendix: conservation derivation}.

\subsubsection*{Subleading Order}

One may also define charges $Q^{\text{in}}[Y^I]$ and $Q^{\text{out}}[Y^I]$ associated to the subleading soft theorem. They are parametrized by an arbitrary vector field $Y^I: \mathbb{R}^9 \rightarrow \mathbb{R}^9$ and enjoy the conservation law
\begin{equation}
    \begin{split}
        \langle \text{out}|\Big(Q^{\text{out}}[Y^I] - Q^{\text{in}}[Y^I]\Big)|\text{in}\rangle = 0.
    \end{split}
\end{equation}
These charges similarly decompose into hard and soft terms
\begin{equation}
    Q^{\text{in}}[Y^I] = Q^{\text{in}}_H[Y^I] + Q^{\text{in}}_S[Y^I],
    \label{eqn: QY}
\end{equation}
which may also be defined via their action on the asymptotic states
\begin{equation}
    \begin{split}
        Q_H^{\text{in}}[Y^I] |\text{in}\rangle &= i \sum_{j \in |\text{in}\rangle} \bigg[Y^I(v_j)\partial_{v_j^I} - \frac{i}{2} \partial_{v_j^J} Y^I(v_j)S_j^{IJ} - \frac{1}{9} \partial_{v_j^I} Y^I(v_j) \omega_j \partial_{\omega_j}\bigg] |\text{in}\rangle \\ Q_S^{\text{in}}[Y^I] |\text{in}\rangle &= - \frac{i}{\kappa} \lim_{\omega_s \rightarrow 0} \bigg[\int d^9 v_s ~ Y^I(v_s) ~ \partial_{v_s^J} \hspace{2pt} \mathbb{P} \hspace{2pt} \widetilde{G}^{(0) \text{in} \dagger}_{IJ}(\omega_s,v_s)\bigg] |\text{in}\rangle,
    \end{split}
\label{eqn:mtheorysubleadingcharge}
\end{equation}
where $\widetilde{G}^{(0) \text{in} \dagger}_{IJ}(\omega_s,v_s)$ is a slightly different linear operator acting on $G^{\text{in} \dagger}_{IJ}(\omega_s,v_s)$, $\mathbb{P} = 1 + \omega_s \partial_{\omega_s}$ is responsible for projecting out the contribution from the leading term in the soft expansion, and $S_j^{IJ}$ is the spin angular momentum acting on the $j^{th}$ particle. The out charges are constructed similarly.

\subsection{Asymptotic Symmetries}
\label{MthSym}

These conserved charges are associated to certain symmetries present in an arbitrary theory of gravity formulated in an asymptotically flat spacetime known as \textit{asymptotic symmetries}. The asymptotic symmetries contain the Poincar\'e group as a subgroup, but have an even richer structure. They are diffeomorphisms that preserve the asymptotic flatness conditions of the metric, $g_{\mu \nu},$ as $r \rightarrow \infty$ for a suitably defined radial coordinate. Such diffeomorphisms are infinitesimally generated by a vector field $\xi^\mu$ which is asymptotically a Killing vector field.

This vector field is parameterized by an arbitrary function of the nine coordinates, $v^I$, and an arbitrary vector field
\footnote{For concreteness, we note that in the coordinates $(u,r,\tilde{v}^a)$, where $u = t - \sqrt{\vec{x} \cdot \vec{x}}$, $ r = \sqrt{\vec{x} \cdot \vec{x}}$, and 
$\tilde{v}^a$ are the coordinates on $S^9$, the most general vector field is given by \cite{Colferai:2020rte}
\begin{equation}
    \xi^\mu \partial_\mu = \Big[f(\tilde{v}) + \frac{u}{9} D_a \tilde{Y}^a(\tilde{v})\Big]\partial_u - \frac{r}{9} D_a \tilde{Y}^a(\tilde{v})\partial_r + \tilde{Y}^a(\tilde{v}) \partial_a + \cdots
    \label{eqn: vector field}
\end{equation}
where $D_a$ is the covariant derivative on the sphere, $\tilde{Y}^a(\tilde{v})$ is related to $Y^I(v)$ through coordinate transformations, and `$\cdots$' includes terms that are subleading order in a large $r$ expansion. 

Although the $\frak{bms}_d$ algebra for $d > 4$ is well understood, the extension of $\tilde{Y}^a(\tilde{v})$ from conformal Killing vectors on $S^{d-2}$ to generators of arbitrary diffeomorphisms of $S^{d-2}$ remains unsettled. Nevertheless, this is often assumed, and we will permit such a general form of $\xi^\mu$ for our purposes of studying M-theory in a flat background.}
\begin{equation}
    \begin{split}
        f&: \mathbb{R}^9 \rightarrow \mathbb{R}\\
        Y^I&: \mathbb{R}^9 \rightarrow \mathbb{R}^9.
    \end{split}
\end{equation}
These are precisely the parameters that appear in the leading and subleading charges. 

Because $f$ and $Y^I$ are arbitrary, these asymptotic symmetries include two functions' worth of degrees of freedom -- we have an infinite-dimensional symmetry group containing Poincar\'e. Though they are diffeomorphisms, because asymptotic symmetries have non-trivial behavior at conformal infinity, they are genuine symmetries of a theory that act non-trivially on its Hilbert space; therefore, they each have a corresponding conserved charge. If we set $Y^I(v)=0$, then we get the so-called \textit{supertranslations}, which are associated to the conserved charge $Q[f]$ and generalize translations. Setting $f(v)=0$ instead gives the so-called \textit{superrotations}, which are associated to the conserved charges $Q[Y^I]$ and generalize rotations and boosts.

\subsubsection*{Poincar\'e Subalgebra}

As these symmetries generalize the Poincar\'e group, it is especially interesting to see how the Poincar\'e algebra is realized. Indeed, the vector field $\xi^\mu \partial_\mu$ of Equation \eqref{eqn: vector field} reduces to generators of global translations ($\xi^\mu \partial_\mu = \partial_\nu$) and Lorentz transformations ($\xi^\mu \partial_\mu = x_\nu \partial_\sigma - x_\sigma \partial_\nu$) for specific choices of $f$ and $Y^I$. 

The conserved charges $Q^{\text{in/out}}[f,Y^I] = Q^{\text{in/out}}[f] + Q^{\text{in/out}}[Y^I]$
also reduce to the Poincar\'e generators for precisely these $f$ and $Y^I.$ Moreover, one can compute the left hand side of
\begin{equation}
    \langle \text{out}| \Big(Q^{\text{out}}[f,Y^I] - Q^{\text{in}}[f,Y^I]\Big)|\text{in}\rangle = 0,
    \label{eqn: M-theory conservation}
\end{equation}
for these choices of $f$ and $Y^I$ and demonstrate that these conservation laws reduce to conservation of momentum $P^\mu$ and angular momentum $J^{\mu \nu}.$ This is summarized in the first three columns of Table \ref{tab: poincare symmetry in BFSS 3}.

\section{Soft Theorems in the BFSS Matrix Model}
\label{sec: soft theorems}

In this section, we discuss the dual description for the leading and subleading soft graviton theorems in the matrix model. We argue that a soft D0-brane bound state (dual to a soft supergraviton) is one that contains $N_s$ D0-branes with $N_s \ll N_j$ where $N_j$ labels the number of D0-branes in the other bound states. Moreover, we show that this soft limit is naturally related to the decompactified M-theory limit. We also demonstrate that the soft expansion in M-theory is recast as a $1/N$ expansion in the matrix quantum mechanics with the leading term scaling $\propto \hspace{2pt} N$ which is responsible for a soft pole in the scattering amplitude. In Appendix \ref{appendix: first principles argument}, we present first-principle evidence that D0-brane scattering actually does obey precisely this soft expansion.

\subsection{Soft Expansion}
\label{BFSSSGT}

Because M-theory scattering amplitudes have a soft expansion of the form given by Equation \eqref{eqn:Mth SGT}, it follows by use of the duality (Equation \eqref{eqn: BFSS duality}), that D0-brane scattering amplitudes in the BFSS matrix model have an analogous soft expansion
\begin{equation}
    \mathcal{A}_{\text{BFSS}}(N_s,v_s^I,\varepsilon; N_j,v_j^I,\Theta_j) = \bigg[\mathcal{S}^{(-1)} + \mathcal{S}^{(0)} + \cdots\bigg] \mathcal{A}_{\text{BFSS}}(N_j,v_j^I,\Theta_j),
    \label{eqn: soft expansion BFSS}
\end{equation}
where $(N_s,v_s^I,\varepsilon)$ are the quantum numbers for the soft D0-brane bound state and $(N_j,v_j^I,\Theta_j)$ schematically denote the quantum numbers for the remaining D0-brane bound states of which there can be several. It remains to find the kinematic regime for which an external D0-brane may be considered `soft.'

\subsection{Soft Limit}

Recall that the soft limit in M-theory is the one for which $\omega_s/\omega_j \rightarrow 0$, where the scales $\omega_s$ and $\omega_j$ are defined as in Equation \eqref{eqn: graviton parametrization}. One can verify that $\omega_s$ and $\omega_j$ are related to the graviton's momentum in the longitudinal direction via $q_s^+ = \omega_s$ and $p_j^+ = \omega_j.$ From the duality dictionary, it follows that the \textit{soft limit} in the matrix model is given by
\begin{equation}
    \textbf{\text{Soft limit:}} \hspace{30pt} \frac{N_s}{N_j} \rightarrow 0 \hspace{30pt} v_s^I,v_j^I = \text{fixed}.
\end{equation}
This implies that the number of D0-branes in a soft D0-brane bound state is dwarfed by the number of D0-branes in the hard particle bound states. The soft expansion, is then an expansion in $N_s/N_j$.

This soft limit is achieved naturally in the M-theory limit provided one holds $N_s$ fixed. This implies that the ratio $N_s/N_j \sim 1/N$, so the soft expansion in gravity gets translated to a $1/N$ expansion in the dual gauge theory with leading term $\mathcal{S}^{(-1)} ~ \propto~  N $ as was originally noticed in \cite{Miller:2022fvc}
\begin{equation}
    \mathcal{S}^{(n)} ~ \propto ~ \Big(\frac{N_s}{N_j}\Big)^n \sim \Big(\frac{1}{N}\Big)^n.
\end{equation}

\subsection{Leading and Subleading Soft Terms}

It remains to find the analogous expressions for $\mathcal{S}^{(-1)}$ and $\mathcal{S}^{(0)}$ in the matrix model. To this end, one may use the duality dictionary to replace the M-theory quantum numbers with the corresponding BFSS quantum numbers. Some simple algebra shows that the leading soft term takes the following form in BFSS (defining $v_{sj} \equiv v_s - v_j$)
\begin{equation}
    \mathcal{S}^{(-1)} = -2 \kappa \sum_{j=1}^n \eta_s \eta_j \frac{N_j}{N_s} \frac{e_{IJ}\hspace{2pt}v_{sj}^I v_{sj}^J}{v_{sj}^2},
    \label{eqn: leading BFSS}
\end{equation}
where we have decomposed the polarization on the basis \eqref{eqn:polbasis}, $\varepsilon^{\mu \nu} = (e^{IJ}) \varepsilon_{IJ}^{\mu \nu}$. One may perform an analogous computation for the subleading soft term. We find\footnote{The derivation is slightly more technical than the leading soft theorem case due to the presence of the 11d orbital $J^{\mu \nu}_j$ and spin $S^{\mu \nu}_j$ angular momentum operators. Fortunately, the authors of \cite{Kapec_2018} have already performed a similar analysis in the context of celestial holography which outlines the procedure.}
\begin{equation}
    \begin{split}
    \mathcal{S}^{(0)} = \frac{\kappa}{2} \sum_{j=1}^n \eta_s \eta_j e^{IJ} \hspace{1pt} \bigg[\mathcal{P}^K_{~ IJ}(v_{sj}) \partial_{v_j^K} + \frac{N_j}{9R}\partial_{v_s^K} \mathcal{P}^K_{~ IJ}(v_{sj}) \partial_{N_j/R} - \frac{i}{2}\partial_{v_s^{[K}}\mathcal{P}^{L]}_{~ IJ}(v_{sj})S_{j,KL}\bigg].
    \end{split}
    \label{eqn: subleading BFSS}
\end{equation}
The spin operator $S^{IJ}_j$ in the matrix model is just the spin part of the BFSS $SO(9)$ generator, $S^{IJ} = (8R)^{-1} \text{Tr}(\Psi^\alpha \Gamma^{[IJ]}_{\alpha \beta} \Psi^\beta)$, acting on the $j^{th}$ particle, and we have defined
\begin{equation}
    \mathcal{P}^K_{~ IJ}(v) = \frac{1}{2}\Big( v_I \delta_J^K + v_J \delta_I^K + \frac{2}{9}v^K \delta_{IJ} - \frac{4}{v^2} v^K v_I v_J \Big).
\end{equation}

\subsection{Evidence for the Soft Theorem}

In this section, we have conjectured an explicit expression for scattering of soft D0-branes in BFSS given by Equation \eqref{eqn: soft expansion BFSS} where the leading and subleading terms in the expansion are provided in Equations \eqref{eqn: leading BFSS} and \eqref{eqn: subleading BFSS}. This conjecture crucially assumes that BFSS amplitudes are related to M-theory amplitudes by the duality dictionary outlined in Section \ref{scatteringduality}. However, this correspondence has only been checked in a very limited setting. As such, it is paramount to give new first-principle evidence that D0-brane scattering amplitudes genuinely obey this soft theorem. 

In Appendix \ref{appendix: first principles argument} we provide a first-principle derivation of the soft theorem. Specifically, we apply the standard soft graviton theorem to M-theory compactified on a spatial circle to derive a soft theorem for D0-branes in type IIA string theory. Afterwards, we take the BFSS limit of the type IIA picture wherein these D0-brane scattering amplitudes become non-relativistic and precisely the BFSS scattering amplitudes by definition; in this limit, one recovers exactly the soft factors of Equations \eqref{eqn: leading BFSS} and \eqref{eqn: subleading BFSS}. This result bolsters the claim that BFSS gives a holographic description of flat space M-theory with duality dictionary given by Table \ref{tab: Kinematics Dictionary}. Moreover, it is a crucial ingredient to demonstrating that BFSS enjoys infinitely many conserved charges associated to asymptotic symmetries as will be discussed in the following section.

\section{Conserved Charges in the BFSS Matrix Model}
\label{sec: conserved charges}

In this section, we will use the soft theorem in the matrix model to construct an infinite set of conserved charges analogous to those charges associated to supertranslation and superrotation symmetry in M-theory. We show that a subset of these charges encode the conservation laws associated to 11d Poincar\'e symmetry and imply, in particular, that scattering amplitudes in the matrix model become Lorentz invariant in the large $N$ limit. This Lorentz symmetry has been long anticipated but previously unverified. Moreover, we show how adding a soft D0-brane to a scattering amplitude can be equivalent to changing the positions and velocities of the hard D0-branes by arbitrary, independent amounts analogous to how a soft graviton can change the position a particle exits on the celestial sphere in M-theory. 

Leveraging the soft theorems to construct an infinite family of conserved charges in the BFSS matrix model follows almost an identical line of reasoning as in the M-theory case. We begin by defining a \textit{creation operator for a soft, incoming D0-brane}, $\mathcal{G}_{IJ}^{\text{in}\dagger}(N_s,v_s^I),$ by its action on asymptotic states. This D0-brane operator should add $N_s$ D0-branes to an asymptotic state where $N_s \sim \mathcal{O}(1)$.\footnote{There is a small, technical subtlety because the total number of D0-branes is fixed by the rank of the gauge group, $N.$ This means that in order to ``create'' D0-branes, one must also remove several D0-branes from one of the other bound states. Fortunately, this is of no consequence in the soft (large $N$) limit because one is only removing an $\mathcal{O}(1)$ number of D0-branes from a bound state which already contains $\mathcal{O}(N)$ D0-branes. In the M-theory limit, one is removing a finite number of D0-branes from a bound state with infinitely many D0-branes, and the effect vanishes.}

By following the derivation that soft theorems imply conservation laws in gravitational theories, one may define conserved charges at leading and subleading order which closely resemble the corresponding conserved charges in M-theory. The main idea is that one can apply a specific integral transform to Equation \eqref{eqn: soft expansion BFSS} which allows one to recast the soft theorem as a Ward identity associated to these conserved charges.  A thorough description of these charges and derivation of their conservation laws is presented in Appendix \ref{appendix: conservation derivation}, and we will summarize the results below.

\subsection{Conserved Charges: Leading Order}
\label{sec: leading charges}

One may define charges $\mathcal{Q}^{\text{in}}[f]$ and $\mathcal{Q}^{\text{out}}[f]$ associated to the leading soft theorem and respectively act on in and out states. The conservation of these charges now reads 
\begin{equation}
    \langle\text{out}|\Big(\mathcal{Q}^{\text{out}}[f] - \mathcal{Q}^{\text{in}}[f]\Big)|\text{in}\rangle = \mathcal{O}\Big(\frac{1}{N}\Big).
\end{equation}
It follows that the leading soft theorem in the matrix model implies the existence of infinitely many conserved charges which are associated to supertranslations in the M-theory picture; however, these charges are only conserved in the M-theory limit! This is a pleasing result because DLCQ breaks the asymptotic symmetries of M-theory, so only in the decompactified M-theory limit do we expect the infinite family of conserved charges to be recovered.

Both the in and the out charge again split up into the sum of a hard and a soft term, $\mathcal{Q}^{\text{in}}[f] = \mathcal{Q}^{\text{in}}_H[f] + \mathcal{Q}^{\text{in}}_S[f]$, which are defined by their action on the asymptotic states,
\begin{equation}
    \begin{split}
        \mathcal{Q}_H^{\text{in}}[f] |\text{in}\rangle &= \sum_{j \in |\text{in}\rangle} \frac{N_j}{R} f(v_j) |\text{in}\rangle \\
        \mathcal{Q}_S^{\text{in}}[f] |\text{in}\rangle &= -\frac{N_s}{4\kappa R} \int d^9 v_s ~ f(v_s) ~ \partial_{v_s^I} \partial_{v_s^J}\widetilde{\mathcal{G}}^{(-1) \text{in} \dagger}_{IJ}(N_s,v_s) |\text{in}\rangle.
        \label{eqn: bfss supertranslation charge definition}
    \end{split}
\end{equation}
Here, $\widetilde{\mathcal{G}}^{(-1) \text{in} \dagger}_{IJ}(N_s,v_s)$ is an integral transformation of $\mathcal{G}^{\text{in} \dagger}_{IJ}(N_s,v_s)$ described in Appendix \ref{appendix: conservation derivation}.

\subsection{Conserved Charges: Subleading Order}
\label{sec: subleading charges}

Next, we define the subleading charges $\mathcal{Q}^{\text{in}}[Y^I]$ and $\mathcal{Q}^{\text{out}}[Y^I]$, whose conservation law reads\footnote{The $\mathcal{O}(1/N)$ term on the right hand side of this set of conservation laws actually decays slightly slower than $1/N$; however, the right hand side still vanishes in the M-theory limit. For technical details see Appendix \ref{appendix: conservation derivation}.}
\begin{equation}
    \langle\text{out}|\Big(\mathcal{Q}^{\text{out}}[Y^I] - \mathcal{Q}^{\text{in}}[Y^I]\Big)|\text{in}\rangle = \mathcal{O}\Big(\frac{1}{N}\Big).
\end{equation}
It follows that the subleading soft theorem in the matrix model implies the existence of infinitely many more conserved charges which are associated to superrotations in the M-theory picture; these charges are also only conserved in the M-theory limit!

As usual, the charges may be decomposed into hard and soft terms, $\mathcal{Q}^{\text{in}}[Y^I] = \mathcal{Q}^{\text{in}}_H[Y^I] + \mathcal{Q}^{\text{in}}_S[Y^I]$ which act on asymptotic states as
\begin{equation}
    \begin{split}
        \mathcal{Q}_H^{\text{in}}[Y^I] |\text{in}\rangle &= i \sum_{j \in |\text{in}\rangle} \bigg[Y^I(v_j)\partial_{v_j^I} - \frac{i}{2} \partial_{v_j^J} Y^I(v_j) S_j^{IJ} - \frac{N_j}{9 R} \partial_{v_j^I} Y^I(v_j)  \partial_{N_j/R}\bigg] |\text{in}\rangle \\ 
        \mathcal{Q}_S^{\text{in}}[Y^I] |\text{in}\rangle &=  - \frac{i}{\kappa} \int d^9 v_s ~ Y^I(v_s) ~ \partial_{v_s^J} \mathbb{P} \hspace{2pt} \widetilde{\mathcal{G}}^{\hspace{1pt}(0) \hspace{1pt} \text{in} \hspace{1pt} \dagger}_{IJ}(N_s,v_s) |\text{in}\rangle.
    \end{split}
    \label{eqn: bfss superrotation charge definition}
\end{equation}
Because $N_j$ is quantized, the `derivative' operator, $\partial_{N_j/R},$ appearing in the hard charge should be approximated by finite differences (this converges to the continuum derivative in the M-theory limit). In addition, $\widetilde{\mathcal{G}}^{\hspace{1pt}(0) \hspace{1pt} \text{in}\hspace{1pt} \dagger}_{IJ}(N_s,v_s)$ is a particular integral transformation of $\mathcal{G}^{\hspace{1pt}\text{in} \hspace{1pt}\dagger}_{IJ}(N_s,v_s)$ and $\mathbb{P}$ is responsible for removing any contributions from the leading soft term. See Appendix \ref{appendix: conservation derivation} for details.

\subsection{11d Poincar\'e Symmetry}
\label{BFSSconservlaw}

In Section \ref{MthSym}, we noted that for specific choices of $f$ and $Y^I$, the conservation laws for the charges, $Q[f]$ and $Q[Y^I]$, associated to the leading and subleading soft graviton theorem in M-theory reduce to the conservation laws implied by Poincar\'e invariance. A similar effect miraculously occurs in the matrix model dual. We begin by examining the conservation law associated to the charges $\mathcal{Q}^{\text{in/out}}[f,Y^I] = \mathcal{Q}^{\text{in/out}}[f] + \mathcal{Q}^{\text{in/out}}[Y^I]$ for these $f$ and $Y^I$
\begin{equation}
    \langle \text{out}| \Big(\mathcal{Q}^{\text{out}}[f,Y^I] - \mathcal{Q}^{\text{in}}[f,Y^I]\Big)|\text{in}\rangle = \mathcal{O}\Big(\frac{1}{N}\Big).
    \label{eqn: BFSS conservation}
\end{equation}
One may use an integration by parts argument to demonstrate that the soft part of the charge $\mathcal{Q}_S^{\text{in/out}}[f,Y^I]$ vanishes identically for the pairs $(f,Y^I)$ discussed in Section \ref{MthSym} and presented in Table \ref{tab: poincare symmetry in BFSS 3}. Moreover, the conservation equation reduces to several known conservation laws for these pairs including: conservation of D0-branes, conservation of momentum, conservation of energy, $SO(9)$ rotational invariance, and Galilean boost invariance. This is also summarized in Table \ref{tab: poincare symmetry in BFSS 3}. These symmetries have been previously studied and are known to exist in the matrix model -- recapturing them provides a nice consistency check of this formalism.

However, the charge conservation equation also implies two new sets of conservation laws which are identified with the conservation of angular momentum $J_{+I}$ and $J_{+-}$ in the M-theory description. Indeed, upon using the duality dictionary (Table \ref{tab: Kinematics Dictionary}), the BFSS conservation laws exactly match the conservation laws associated to Poincar\'e symmetry in M-theory. For example, plugging $f(v) = 0$ and $Y^K(v) = v^K$ into the M-theory conserved charge $Q[f,Y^I]$ yields the $J_{+-}$ conservation law
\begin{equation}
    i\mathlarger{\sum}_{j} \hspace{3pt} \eta_j  (v_j^I\partial_{v_j^I} - \omega_j \partial_{\omega_j}) \mathcal{A}_{\text{M}} = 0.
\end{equation}
Doing the same for the BFSS charge $\mathcal{Q}[f,Y^I]$ yields the following conservation law, which we interpret as implying the analog statement of $J_{+-}$ invariance in the large $N$ limit of BFSS 
\begin{equation}
    i\mathlarger{\sum}_{j} \hspace{3pt} \eta_j  (v_j^I\partial_{v_j^I} - \mathlarger{\frac{N_j}{R}} \partial_{N_j/R}) \mathcal{A}_{\text{BFSS}} = \mathcal{O}\Big(\frac{1}{N}\Big),
\end{equation}
where we recall that the supergraviton energy scale $\omega_j$ in M-theory is identified with the number of D0-branes $N_j/R$ in BFSS. Explicit expressions for conservation laws associated to the other Poincar\'e generators are given in Appendix \ref{appendix: Poincare symmetry}. Altogether, we have demonstrated that the soft theorems in the matrix model imply that D0-brane scattering amplitudes in BFSS enjoy a full 11d Poincar\'e invariance which is only realized in the $N \rightarrow \infty$ limit!

\begin{table}[H]
\centering
\begin{tabular}{||c||c||c||c||c||} 
\hhline{|t:=:t:=:t:=:t:=:t|}
 \rule{0pt}{2ex} $f$                 & $Y^K\partial_K$        &  M-theory & BFSS  \\[1ex]
\hhline{|:=::=::=::=:|}
  \rule{0pt}{5ex} $1$ &  0      & $P^+$ & Number of D0-branes     \\[2ex]
\hline
\rule{0pt}{5ex} $\sqrt{2} \hspace{2pt} v^I$ & 0       & $P^I$ & Momentum     \\[2ex]
\hline
\rule{0pt}{5ex} \rule{0pt}{5ex} $v^{\hspace{1pt}2}$ & 0       &  $P^-$ & Energy     \\[2ex]
\hhline{|:=::=::=::=:|}
\rule{0pt}{5ex} $0$ & $-(v_I \partial_J - v_J \partial_I)$     & $J_{IJ}$   & $SO(9)$ Invariance      \\[2ex]
\hline
\rule{0pt}{5ex} $0$ & $\frac{1}{\sqrt{2}}\partial_I$      &  $J_{-I}$  & Galilean Boost Invariance     \\[2ex]
\hline
\rule{0pt}{5ex} $0$ & $\frac{1}{\sqrt{2}}(v^2 \partial_I-2v_I v^K \partial_K)$   & $J_{+I}$   &  New Symmetry     \\[2ex]
\hline
\rule{0pt}{5ex} $0$ & $v^K\partial_K$        & $J_{+-}$    &  New Symmetry    \\[2ex]
\hhline{|b:=:b:=:b:=:b:=:b|}
\end{tabular}
\caption{Conservation laws associated to various choices of $f$ and $Y^I$ in the M-theory description and BFSS dual. The first five rows of these conservation laws reduce to the known global symmetries of BFSS; however, the last two are new symmetries. The full set is equivalent to Poincar\'e symmetry in the M-theory picture.}
\label{tab: poincare symmetry in BFSS 3}
\end{table}

\subsection{Infinitely Many New Symmetries}

Much like translations and rotations can be promoted to supertranslations and superrotations in M-theory, for a general function $f$ and vector field $Y^I$, the conserved charges $\mathcal{Q}^{\text{in/out}}[f]$ and $\mathcal{Q}^{\text{in/out}}[Y ^I]$ generate an infinite dimensional symmetry group that can move individual D0-brane bound states in arbitrarily different ways.

Consider a situation where there are $n$ asymptotic D0-brane bound states. If we view these bound states as being point-like objects travelling in the nine transverse directions with positions $x_j \in \mathbb{R}^9$ and velocities $v_j \in \mathbb{R}^9$, the asymptotic states are Gaussian wavepackets whose location in position space at some time is predominantly supported in the disjoint regions $\Omega_1,...,\Omega_n \subset \mathbb{R}^9$ (with $\Omega_j$ centered around $x_j$) and whose velocities are predominantly supported in the disjoint regions $\widetilde{\Omega}_1,...,\widetilde{\Omega}_n \subset \mathbb{R}^9$ (with $\widetilde{\Omega}_j$ centered around $v_j$).

\begin{figure}[H]
    \centering
    \includegraphics[width=\textwidth]{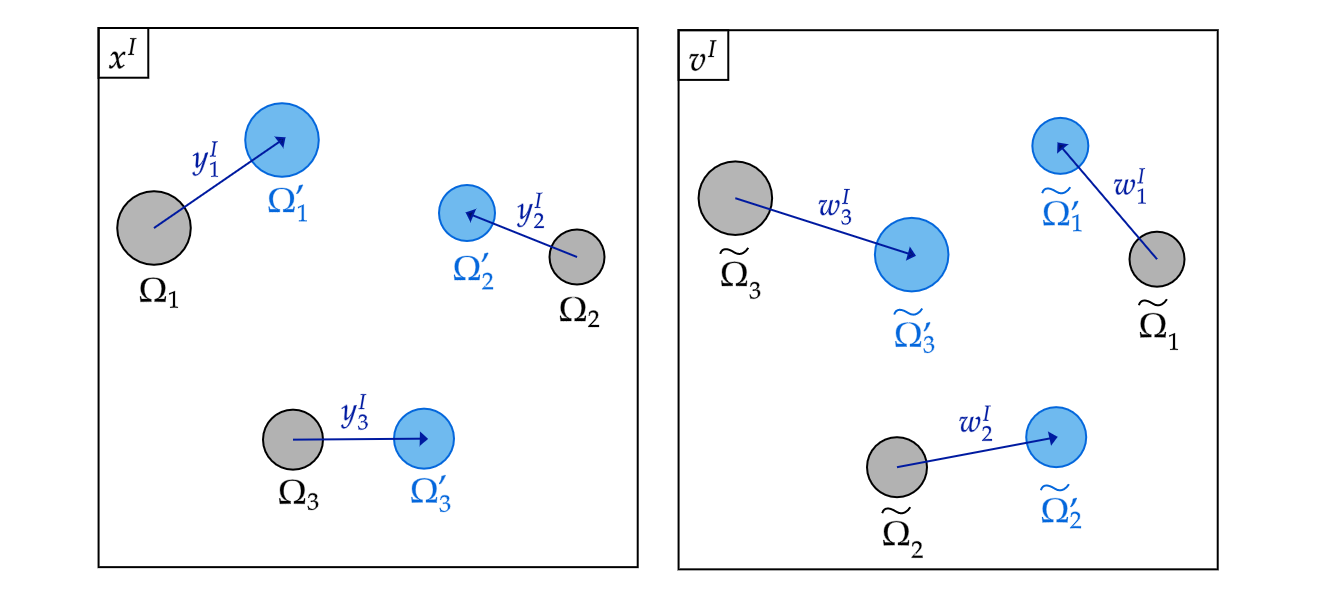}
    \caption{(Left) Action of the unitary operator $\mathcal{U}_1[f]$ which translates the wavepackets by an amount $y_j$ and so shifts their domain of central support in position space according to $\Omega_j \mapsto \Omega_j + y_j$. (Right) Action of the unitary operator $\mathcal{U}_1[Y^I]$ which boosts the wavepackets by an amount $w_j$ and so shifts their domain of central support in momentum space according to $\widetilde{\Omega}_j \mapsto \widetilde{\Omega}_j + w_j$.}
    \label{fig: wavepackets}
\end{figure}

Because $\widetilde{\Omega}_j$ are disjoint, we may choose a function $f(v)$ that is linear on the domains $\widetilde{\Omega}_j$
\begin{equation}
    f(v)|_{\widetilde{\Omega}_j} = y_j \cdot v,
\end{equation}
where $y_1,...,y_n \in \mathbb{R}^9$ are an arbitrary set of vectors. When $y_j$ are all the same, we may simply take $f(v)$ to be the globally-defined linear function $f(v) = y \cdot v$. According to Table \ref{tab: poincare symmetry in BFSS 3}, the corresponding charge is just the BFSS momentum operator $y \cdot P$ which infinitesimally translates all particles in the $y$ direction by the same amount. When the constants $y_j$ differ, however, the wavepackets get infinitesimally translated in different directions. For example, if we define the unitary operator, $\mathcal{U}_a[f],$ generated by the conserved charge $\mathcal{Q}[f]$ as 
\begin{equation}
    \mathcal{U}_a[f] = \exp(i a \hspace{1pt} \mathcal{Q}[f]) = \exp(i a \hspace{1pt} \mathcal{Q}_S[f] + i a \hspace{1pt} \mathcal{Q}_H[f]),
\end{equation}
then we observe that the hard term in $\mathcal{U}_a[f]$ translates the $j^{\text{th}}$ particle according to $x_j \mapsto x_j + a y_j$. Meanwhile, the soft term -- being proportional to a soft D0-brane creation operator -- is responsible for creating a coherent state of soft D0-branes. Thus, we see how adding soft D0-branes can be secretly equivalent to translating the hard D0-branes.

Similarly, we may choose the function $Y^I(v)$ such that it is constant on the domains $\widetilde{\Omega}_j$
\begin{equation}
    Y^I(v)|_{\widetilde{\Omega}_j} = w_j^I,
\end{equation}
where $w_1,...,w_n \in \mathbb{R}^9$ are another arbitrary set of vectors. When $w_j$ are all the same, we may take $Y^I(v)$ to be the globally defined linear function $Y^I(v) = w^I$. According to Table \ref{tab: poincare symmetry in BFSS 3}, the corresponding charge is just the generator of Galilean boosts in the matrix model $w \cdot \mathcal{K}$ (see Section \ref{sec: construction of Poincare generators}) which infinitesimally changes the velocity of particles by an amount $w$. When the constants $w_j$ differ, however, the wavepackets get infinitesimally boosted by different amounts. If we define the unitary operator, $\mathcal{U}_a[Y^I],$ generated by the conserved charge $\mathcal{Q}[Y^I]$ as 
\begin{equation}
    \mathcal{U}_a[Y^I] = \exp(i a \hspace{1pt} \mathcal{Q}[Y^I]) = \exp(i a \hspace{1pt} \mathcal{Q}_S[Y^I] + i a \hspace{1pt} \mathcal{Q}_H[Y^I]),
\end{equation}
then we observe that the hard term in $\mathcal{U}_a[f]$ changes the velocity of the $j^{\text{th}}$ particle according to $v_j \mapsto v_j + a w_j$. The soft term is again responsible for creating a coherent state of soft D0-branes. Thus, we see how adding soft D0-branes can also be secretly equivalent to individually boosting the hard D0-branes by specified amounts. This is summarized in Figure \ref{fig: wavepackets}.

Of course, one could choose even more complicated functions $f(v)$ and $Y^I(v)$ which will distort wavepackets of D0-branes in arbitrarily non-trivial ways. The cost of doing so is integrating the soft D0-brane creation operator against these functions and then including this particle as an asymptotic state in a scattering amplitude.

\section{Asymptotic Symmetries in the BFSS Matrix Model}
\label{sec: asymptotic symmetries}

In this section, we will interpret the aforementioned conservation laws in terms of asymptotic symmetries. Our main finding is that the asymptotic symmetries associated to the 11d large diffeomorphisms of M-theory are manifested in the D0-brane quantum mechanics via the asymptotic symmetries of the metric and RR 1-form gauge field in type IIA string theory. We offer several complementary perspectives and non-trivial consistency checks of this conclusion. Along the way, we find explicit expressions for the hard charges $\mathcal{Q}_H$ in terms of $N \times N$ matrix degrees of freedom.

\subsection{Coupling BFSS to a Background 11d Metric}
\label{background field AGS}

BFSS gives a dual description of M-theory with flat space metric, $\eta_{\mu \nu}$. We begin by briefly reviewing how the BFSS matrix model is deformed when one instead considers M-theory with a weakly curved background 11d metric, $\eta_{\mu \nu} + g_{\mu \nu}$, where $g_{\mu \nu}$ is independent of the coordinate $x^-$ parametrizing the compact direction. In a series of papers on D0-brane quantum mechanics, Taylor and Van Raamsdonk have argued that in the presence of such a background field, the BFSS action gets modified by the following interaction term \cite{Taylor:1998tv, Taylor:1999gq}\footnote{In this context, the time $t$ appearing in the matrix theory action is identified with the M-theory time $x^+$ from lightcone quantization.}
\begin{equation}
    S_{\text{int}}[g_{\mu \nu}] = \frac{1}{2} \int dt ~ \sum_{m=0}^\infty \frac{1}{m!} \big(\partial_{I_1} \cdots \partial_{I_m} g_{\mu \nu}\big) T^{\mu \nu(I_1 \cdots I_m)}.
\end{equation}
Here, $\partial_{I_1} \cdots \partial_{I_m} g_{\mu \nu}$ is a time-dependent coupling constant constructed from the background field, $g_{\mu \nu}$, and does not depend on the $N \times N$ matrix degrees of freedom. In particular, it is defined as the Taylor expansion of the 11d fields about the spatial origin in the transverse directions $x^I = 0$
\begin{equation}
    \partial_{I_1} \cdots \partial_{I_m} g_{\mu \nu}(t) = \frac{\partial}{\partial x^{I_1}} \cdots \frac{\partial}{\partial x^{I_m}} g_{\mu \nu}(x^+, x^I)\Big|_{x^I = 0, ~ x^+ = t}.
\end{equation}

By contrast, $T^{\mu \nu}$ is the \textit{matrix theory stress-energy tensor} and $T^{\mu \nu(I_1 \cdots I_m)}$ are its \textit{multipole moments.} They are built out of the $N \times N$ matrix degrees of freedom, but have no background dependence. Furthermore, the stress-energy tensor satisfies the following conservation equation \cite{VanRaamsdonk:1997in, Taylor:1998tv, Taylor:1999gq}
\begin{equation}
    \partial_0 T^{+\nu(I_1 \cdots I_m)} = T^{I_1 \nu(I_2 \cdots I_m)} + \cdots + T^{I_m \nu(I_1 \cdots I_{m-1})}.
\end{equation}
We will need explicit expressions for a few components of the stress-energy tensor. In particular
\begin{equation}
    \begin{split}
        T^{++} &= \frac{1}{R}\hspace{2pt}\text{STr}\big(\mathds{1}\big) \\
        T^{+I} &= \frac{1}{R}\hspace{2pt}\text{STr}\big(\partial_t X^I\big) \\
        T^{+-} &= \frac{1}{R}\hspace{2pt}\text{STr}\Big(\frac{1}{2} \partial_t X^I \partial_t X^I - \frac{1}{4}[X^I,X^J][X^I,X^J] + \frac{1}{2}\Psi^\alpha \Gamma^I_{\alpha \beta}[X^I,\Psi^\beta]\Big),
        \label{eqn: stress-energy tensor}
    \end{split}
\end{equation}
where ``STr'' denotes the symmetrized trace of a product of matrices. The multipole moments of the matrix model stress-energy tensor are related to the zeroth order moments according to
\begin{equation}
    T^{\mu \nu(I_1 \cdots I_m)} = \text{Sym}(T^{\mu \nu}; X^{I_1}, \cdots, X^{I_m}) + T^{\mu \nu(I_1 \cdots I_m)}_{\text{fermion}},
\end{equation}
where the first term denotes that we should insert the bosonic matrices, $X^I$, in the symmetrized traces considered in Equation \eqref{eqn: stress-energy tensor} and the second term gives additional contributions to the multiple moments containing the fermions, $\Psi^\alpha.$ In particular, we will need the expressions
\begin{equation}
    \begin{split}
        T^{++(J)}_{\text{fermion.}} &= 0 \\
        T^{+I(J)}_{\text{fermion.}} &= \frac{1}{8R}\text{Tr}\big(\Psi^\alpha \Gamma^{[IJ]}_{\alpha \beta} \Psi^\beta\big) \\
        T^{+-(J)}_{\text{fermion}} &= \frac{1}{16R} \text{Tr}\big(\Psi^\alpha [X^K,X^L]\Gamma^{[KLJ]}_{\alpha \beta} \Psi^\beta + 2i \Psi^\alpha \partial_t X^K \Gamma^{[KJ]}_{\alpha \beta} \Psi^\beta \big).
    \end{split}
\end{equation}

\subsection{Asymptotic Symmetries in BFSS}
\label{sec: hard charge construction}

Now, we will consider a setup where the 11d metric is pure gauge and related to the flat metric by an infinitesimal diffeomorphism generated by the vector field $\xi_\mu$
\begin{equation}
    g_{\mu \nu} = \partial_\mu \xi_\nu + \partial_\nu \xi_\mu.
\end{equation}
Moreover, we will study the situation where the gauge parameter $\xi_\mu$ is non-vanishing on $\mathcal{I}^\pm$ and is given by an expression of the form \eqref{eqn: vector field}. It is known that in quantum field theory, a large diffeomorphism like this only affects one's action up to a possible boundary term. Such a boundary term is precisely the conserved charge $Q[\xi_\mu]$ described in Equations \eqref{eqn:mtheoryleadingcharge} and \eqref{eqn:mtheorysubleadingcharge}.

In the case of D0-brane quantum mechanics, the same effect miraculously appears. We begin by considering $S_{\text{int}}[\partial_\mu \xi_\nu + \partial_\nu \xi_\mu]$. One may use the conservation equation to integrate this expression by parts. The resulting action can be written as the integral of a total derivative; therefore, it reduces to the following boundary term
\begin{equation}
    S_{\text{int}}[\partial_\mu \xi_\nu + \partial_\nu \xi_\mu] =  \sum_{m=0}^\infty \frac{1}{m!} \big(\partial_{I_1} \cdots \partial_{I_m} \xi_\mu) T^{+\mu(I_1 \cdots I_n)}\bigg|_{-\infty}^{+\infty}.
\end{equation}
As such, we may define the candidate incoming conserved charge, $\mathcal{Q}_H^{\text{in}}[\xi_\mu]$, by
\begin{equation}
    \mathcal{Q}_{H}^{\text{in}}[\xi_\mu] = \sum_{m=0}^\infty \frac{1}{m!} \big(\partial_{I_1} \cdots \partial_{I_m} \xi_\mu) T^{+\mu(I_1 \cdots I_n)}\bigg|_{-\infty}.
    \label{eqn: hard charge}
\end{equation}
The candidate outgoing conserved charge is defined analogously. One can only recover the hard part of the conserved charge because $g_{\mu \nu}$ is a non-dynamical background field in this formalism.

\subsection{Relation to RR 1-form and 10d Metric}
\label{sec:backgroundfield}

In \cite{Miller:2022fvc}, we have argued that the soft theorem in the BFSS matrix model may be understood from the lens of studying asymptotic symmetries of the background RR 1-form gauge field, $A_\mu,$ in type IIA string theory.\footnote{In this section, the Greek indices $\mu,\nu$ will also be used to describe the 10d parameters in type IIA. Whether $\mu, \nu = 0,...,10$ or $\mu,\nu = 0,...,9$ will be clear from context.} This was originally noticed because the term $N_j/R$ appearing in the soft theorem is exactly the $U(1)$ charge of a stack of D0-branes under $A_\mu.$\footnote{Note that a stack of D0-branes is BPS, so its $U(1)$ charge is also its mass; therefore, one might guess that asymptotic symmetries of the 10d metric, $h_{\mu \nu}$, in IIA play an equally important role.} 

In the previous section, we have argued that the asymptotic symmetries of M-theory are related to large diffeomorphisms of the 11d graviton field $g_{\mu \nu}$. These two perspectives reconcile because the 11d graviton field is broken down to a 10d metric perturbation, $h_{\mu \nu}$, a RR 1-form gauge field, $A_{\mu},$ and a dilaton, $\phi$ when compactified -- thus, asymptotic symmetries of M-theory are at least partially manifested in the asymptotic symmetries of $A_\mu$. The \textit{full} 11d gravitational asymptotic symmetry group of M-theory is manifested in D0-brane quantum mechanics dual via the asymptotic symmetries of $h_{\mu \nu}$ and $A_\mu$.\footnote{The 10d dilaton doesn't enjoy a standard set of asymptotic symmetries and won't play a role in our story. See \cite{Kapec:2022axw, Cheung:2021yog} for details.}

To see this, we note that the discrete lighcone quantization of M-theory may be viewed as a particular limit of so-called $\widehat{\text{M}}$-theory, which is just ordinary M-theory compactified along the $x^{10}$ direction with radius of compactification $R_c.$ $\widehat{\text{M}}$-theory is related to the DLCQ of M-theory via a large boost in the $x^{10}$ direction with parameter $\gamma = \sqrt{R^2/2R^2_c + 1}$ followed by taking the $R_c \rightarrow 0$ limit \cite{Seiberg_1997}. The background metric perturbation in these two theories are similarly related by a large boost, so the Lorentz transformation with parameter $\gamma$ maps $g_{\mu \nu} \mapsto \widehat{g}_{\mu \nu}$ and also $\xi_\mu \mapsto \widehat{\xi}_\mu$ relating the 11d gauge parameters for the background metric in the two descriptions. The $R_c \rightarrow 0$ limit of $\widehat{\text{M}}$-theory is just type IIA string theory with background fields \cite{Taylor:1999gq}
\begin{equation}
    \begin{split}
        h_{\mu \nu} &= \widehat{g}_{\mu \nu} + \frac{1}{2}\widehat{g}_{10 \hspace{1pt} 10} \hspace{30pt} A_{\mu} = \widehat{g}_{10 \hspace{1pt} \mu} \hspace{30pt} \phi = \frac{3}{4}\widehat{g}_{10 \hspace{1pt} 10}.
        \label{eqn: background field relation}
    \end{split}
\end{equation}
Plugging in the expressions for $\widehat{g}_{\mu \nu}$ in terms of $g_{\mu \nu}$ yields expressions for the type IIA background field in terms of the original 11d metric (see \cite{Taylor:1999gq})

If $g_{\mu \nu}$ is pure gauge, then $\widehat{g}_{\mu \nu}$ is similarly pure gauge, and the 10-dimensional type IIA fields will necessarily be pure gauge as well
\begin{equation}
    h_{\mu \nu} = \partial_\mu \zeta_\nu + \partial_\nu \zeta_\mu \hspace{30pt} A_\mu = \partial_\mu \lambda,
\end{equation}
for some 10d gauge parameters $(\zeta_\mu,\lambda)$. One can read off expressions for these gauge parameters in terms of the 11d gauge parameter, $\widehat{\xi}_\mu$ by plugging $\widehat{g}_{\mu \nu} = \partial_\mu \widehat{\xi}_\nu + \partial_\nu \widehat{\xi}_\mu$ into Equation \eqref{eqn: background field relation}.\footnote{Defining the 10d background field $A_\mu$ as in Equation \eqref{eqn: background field relation} yields a function which may also depend on the compact direction $A_{\mu}(t,x^I,x^{10})$. However, these are supposed to be background fields in 10-dimensional type IIA string theory. The correct prescription to remove the $x^{10}$ dependence is to just average over the compactified direction $A_{\mu}(t,x^I) = \int dx^{10} \hspace{2pt} \widehat{g}_{10\hspace{1pt} \mu}(t,x^I,x^{10})$. This implies, in particular, that we may set $\partial_{10} \hspace{1pt} \xi_\mu = 0$ because $\int dx^{10} \hspace{2pt} \partial_{10} \hspace{1pt} \xi_\mu = 0.$}
\begin{equation}
    \begin{split}
        \zeta_0 &= \widehat{\xi}_0 = \frac{1}{\alpha \sqrt{2}} \xi_+ - \frac{\alpha}{\sqrt{2}} \xi_- \\
        \zeta_I &= \widehat{\xi}_I = \xi_I \color{white}\bigg|\color{black} \\ 
        \lambda &= \widehat{\xi}_{10} = \frac{1}{\alpha \sqrt{2}} \xi_+ - \frac{\alpha}{\sqrt{2}} \xi_-  ~,
    \end{split} 
    \label{eqn: relation of gauge params}
\end{equation}
where we have also related the 11d gauge parameter of $\widehat{\text{M}}$-theory, $\widehat{\xi}_\mu$, to the 11d gauge parameter of M-theory, $\xi_\mu$, via the aforementioned Lorentz boost. This expression depends on the parameter $\alpha \approx R_s/\sqrt{2}R \rightarrow 0$. Though several terms in Equation \eqref{eqn: relation of gauge params} appear to be either vanishing or divergent in the $\alpha \rightarrow 0$ limit, they always appear in finite combinations due to an additional rescaling of parameters relating type IIA string theory to the BFSS action (see \cite{Taylor:1999gq,Seiberg_1997}). 

In \cite{Miller:2022fvc}, we have shown that one may construct a hard charge associated to the 10d gauge field, $\mathcal{Q}_H[\lambda]$, via a similar integration by parts procedure as the one outlined in Section \ref{sec: hard charge construction}. Moreover, this charge reproduces Equation \eqref{eqn: bfss supertranslation charge definition} when asymptotic states assume the form given by Equation \eqref{eqn: asymptotic states}. One can construct a similar charge for the 10d metric, $\mathcal{Q}^{\text{in}}_H[\zeta^\mu].$ This gives additional evidence that the asymptotic symmetries of the 11d metric are encoded in the D0-brane quantum mechanics picture through the asymptotic symmetries of $h_{\mu \nu}$ and $A_\mu$ in type IIA string theory and further bolsters the claim that $\mathcal{Q}^{\text{in}}_H[\xi^\mu]$ defined in Equation \eqref{eqn: hard charge} is the correct definition for the hard charge in the matrix model. In \cite{Miller:2022fvc}, we also showed how adding a soft D0-brane can be viewed as turning on a background, pure gauge $A_{\mu}$ field -- this formalism makes clear that it should be interpreted as turning on a background, pure gauge $g_{\mu \nu}$ field which decomposes into pure gauge IIA $A_\mu$ and $h_{\mu \nu}$ fields according to Equation \eqref{eqn: relation of gauge params}.

\subsection{Construction of Poincar\'e Generators in the Matrix Model}
\label{sec: construction of Poincare generators}

\begin{table}
\centering
\begin{tabular}{||c||c||c||} 
\hhline{|t:=:t:=:t:=:t|}
$\xi_\mu \partial^\mu$ & M-theory: $Q_H[\xi^\mu]$ &  BFSS: $\mathcal{Q}_H[\xi^\mu]$ \\ 
\hhline{|:=::=::=:|}
$\partial^+$ & $P^+$ &  $T^{++} = \mathlarger{\frac{N}{R}}$ \\
$\partial^I$ & $P^I$ &  $T^{+I} = \text{Tr}(P^I)$ \\
$\partial^-$ & $P^-$ &  $T^{+-} = H_\text{BFSS}$ \\
\hhline{|:=::=::=:|}
$x^I \partial^J - x^J \partial^I$ & $J^{IJ}$ &  $T^{+J(I)} - T^{+I(J)} = \mathcal{J}^{IJ}$\\
$x^+ \partial^I - x^I \partial^+$ & $J^{+I}$ &  $- T^{++(I)} = \mathcal{K}^I$  \\
$x^- \partial^I - x^I \partial^-$ & $J^{-I}$ & - \\
$x^- \partial^+ - x^+ \partial^-$ & $J^{-+}$ & - \\
\hhline{|b:=:b:=:b:=:b|}
\end{tabular}
\caption{Correspondence between M-theory Poincar\'e generators and the dual BFSS generators which are defined by $\mathcal{Q}_H[\xi^\mu]$ for various choices of $\xi^\mu.$}
\label{tab: correspondence of generators}
\end{table}

Equation \eqref{eqn: hard charge} gives an explicit candidate for the hard charge, $\mathcal{Q}_H[\xi^\mu]$, in the BFSS matrix model associated to a large diffeomorphism in M-theory generated by the vector field $\xi^\mu \partial_\mu.$ As such, it is especially interesting to try to use $\mathcal{Q}_H[\xi^\mu]$ to reproduce the generators of global translations and Lorentz transformations in the matrix model in an effort to identify how the 11d Poincar\'e algebra is actually manifested in terms of the $N \times N$ matrix degrees of freedom.

In the M-theory picture, the conserved charges associated to global translations are the momentum operators $P^+$, $P^I$, and $P^-$. In the BFSS picture, we expect the corresponding charges to be $N/R$ (due to the fact that we are working in a sector of M-theory where $P^+ = N/R$), the BFSS momentum operator $\text{Tr}(P^I)$, and the BFSS Hamiltonian which generates time translations (and so plays a similar role to $P^-$ in the lightcone quantization of M-theory). We also expect the M-theory Lorentz generator $J^{IJ}$ to be dual to the generator of $SO(9)$ rotations in the matrix model
\begin{equation}
    \mathcal{J}^{IJ} = \text{Tr}\big(P^I X^J-P^J X^I \big) + \frac{1}{8R}\text{Tr} \big(\Psi^\alpha \Gamma^{[IJ]}_{\alpha \beta} \Psi^\beta\big),
\end{equation}
while the Lorentz generator $J^{+I}$ should be dual to the generator of Galilean boosts\footnote{Previous expressions for the $SO(9)$ rotation generator and Galilean boost generator in terms of the $N \times N$ matrices are known (see \cite{Lowe:1998wu,deWit:1989vb}); however, it is interesting to see them derived from a completely different perspective.} 
\begin{equation}
    \mathcal{K}^I = \mathlarger{\frac{1}{R}} \text{Tr}(X^I).
\end{equation}
 It is encouraging to find that by plugging in appropriate choices of the vector field $\xi^\mu$ into $\mathcal{Q}_H[\xi^\mu]$, these expectations are exactly realized (see Table \ref{tab: correspondence of generators})!

Unfortunately, one cannot construct the dual description of the M-theory generators $J^{-I}$ and $J^{+-}$ using this technology. This is due to the fact that the background field formalism we have been employing is incompatible with vector fields $\xi^\mu \partial_\mu$ which vary non-trivially in the compact $x^-$ direction \cite{Taylor:1998tv, Taylor:1999gq}. We hope that a suitable modification of the above prescription will allow one to fully construct the $11d$ Lorentz generators using only the $N \times N$ matrix degrees of freedom.

Nevertheless, the correspondence between the symmetry generators of M-theory and those of BFSS afforded by this formalism is a heartening result and offers a non-trivial consistency check regarding how the asymptotic symmetries of M-theory are manifested in the matrix model dual.

\section{Discussion}

In this work, we examined how the soft theorems and infinitely many conserved charges associated to asymptotic symmetries in M-theory are manifested in the matrix model dual. In particular, we made a conjecture about the structure of scattering amplitudes among D0-brane bound states in the limit where the number of D0-branes in one particular bound state is dwarfed by the number of D0-branes in all others. While this conjecture was based on the original description of the BFSS duality, we presented a first-principle calculation of it by considering the BFSS scaling limit of D0-brane scattering amplitudes in type IIA string theory, thus providing a novel test of the BFSS duality in a completely new setting. We showed how this soft theorem implies that BFSS enjoys conservation laws associated to the counterpart of 11d supertranslation and superrotation symmetry in the M-theory picture -- these conservation laws are only realized in the large $N$ limit of the matrix model, which is necessary for consistency with the gravity description. Moreover, we interpreted the infinite dimensional symmetry algebra of BFSS, which gives rise to these conservation laws, in terms of the asymptotic symmetries of the metric and RR 1-form gauge field of type IIA string theory, which appear as background fields. Finally, we showcased how these perspectives reconcile to provide a consistent understanding of 11d Lorentz symmetry in the matrix model. In particular, we have shown that BFSS scattering amplitudes exhibit the full 11d Poincar\'e symmetry of M-theory which emerges only in the large $N$ limit, a claim which has been long anticipated but never verified.

We hope that this project will also springboard future studies of the matrix model and the relationship between celestial holography and BFSS. Celestial holography provides a natural framework for studying the symmetries of flat space, and it would be interesting to see how these ideas are manifested in the matrix model \cite{Pano:2023slc, Donnay:2018neh,Pate:2019mfs,Guevara:2019ypd,Donnay:2020guq,Fotopoulos:2020bqj,Pasterski:2021fjn,Donnay:2022sdg,Kapec:2022hih,Dumitrescu:2015fej,Lysov:2015jrs}. For example, a $w_{1+\infty}$ holographic symmetry algebra associated to soft gravitons was recently identified \cite{Strominger:2021lvk,Guevara:2021abz,Guevara:2022qnm,Himwich:2021dau,Jiang:2021csc,Jiang:2021ovh,Adamo:2021lrv,Bu:2022iak,Schwarz:2022dqf,Mago:2021wje,Ball:2021tmb}. It is known that the $SU(N) \subset U(N)$ part of the BFSS symmetry algebra tends to the $w_{1+\infty}$ algebra as $N \rightarrow \infty$ \red{\cite{Guevara:2022qnm,deWit:1988wri}}. It would be exciting to see if this concordance actually encodes a deep connection. 

It would also be intriguing to investigate how these concepts are related to black holes in the matrix model. In particular, the soft theorem describes how an $\mathcal{O}(1)$ number of D0-branes can get dislodged in scattering amplitudes. This is remniscent of Hawking radiation wherein individual D0-branes are the Hawking quanta of a Schwartzchild black hole, which is a Boltzmann gas of $N$ D0-branes \cite{Klebanov:1997kv,Banks:1997hz,Banks:1997tn,Horowitz:1997fr}. For additional work relating BFSS to black holes, see \cite{Hanada:2016zxj,Hyakutake:2014maa,Hanada:2013rga,Maldacena:2018vsr,Pateloudis:2022ijr,Berkowitz:2016jlq, Lin:2023owt,Anagnostopoulos:2007fw,Catterall:2009xn,Biggs:2023sqw,Maldacena:2023acv}.

Finally, this work leaves a number of interesting follow-up questions and calculations. For example, one might hope to give a worldline proof of the soft theorem using the matrix model Hamiltonian \eqref{eqn: BFSS Hamiltonian}. Recovering the leading term $\propto \hspace{2pt} N$ would complement the derivation provided in Appendix \ref{appendix: first principles argument} and be a very interesting computation in its own right. One could also try to understand how the soft theorems associated to the gravitino and 3-form gauge field in 11d $\mathcal{N} = 1$ supergravity are realized in BFSS. In this work, we have only discussed soft theorems for D0-branes in the matrix model whose polarization tensors correspond to gravitons in the M-theory description -- examining soft theorems for the other particles in the supermultiplet will likely shed light on new symmetries of the matrix model which are only realized in the large $N$ limit. A better understanding of the compact direction and how one could use the formalism of Section \ref{sec: asymptotic symmetries} to construct explicit expressions for the complete set of Poincar\'e generators would also be desirable.

\section*{Acknowledgements}

We would like to thank Daniel Harlow, Daniel Jafferis, Antal Jevicki, Daniel Kapec, Hong Liu, David Lowe, Noah Miller, Atul Sharma, Washington Taylor, Nicolas Valdes, Akshay Yelleshpur Srikant, and especially Andy Strominger for stimulating discussions. We would also like to thank Daniel Harlow, Hong Liu, Noah Miller, Atul Sharma, and Andy Strominger for useful comments on the draft. AT gratefully acknowledges support from NSF GRFP grant DGE1745303.

\appendix

\section{A First-Principle Argument for the Soft Theorem in BFSS}
\label{appendix: first principles argument}

In this Appendix, we provide a derivation that non-relativistic D0-brane scattering amplitudes obey the soft theorem given by Equation \eqref{eqn: soft expansion BFSS}.

\subsubsection*{Soft Theorems and $S^1$ Compactification}

We will begin by revisiting the soft graviton theorem in M-theory. This time, however, we will not consider the DLCQ of M-theory -- rather, we will work with the more familiar quantization on hypersurfaces of constant time. The leading soft graviton theorem obeys the usual universal expression in the limit $\omega_s /\omega_j \rightarrow 0$ where the supergraviton momenta are parametrized by \eqref{eqn: graviton parametrization}
\begin{equation}
     \mathcal{A}_{\text{M}}(q_s^\mu,\varepsilon;p_1^\mu,...,p_n^\mu) = \Big[\mathcal{S}^{(-1)} + \cdots \Big]\mathcal{A}_{\text{M}}(p_1^\mu,...,p_n^\mu),
\end{equation}
where
\begin{equation}
    \mathcal{S}^{(-1)} = \frac{\kappa}{2} \hspace{2pt} \varepsilon_{\mu \nu} \sum_{j=1}^n \eta_s \eta_j \frac{p_j^\mu p_j^\nu}{q_s \cdot p_j}.
    \label{eqn: leading soft compactification}
\end{equation}
For the moment, we will only focus on the leading term in the soft expansion. 

Next, we compactify on a spatial circle with radius of compactification $R_c$ (in the $x^{10}$ direction, for simplicity). Now the momentum in the compact direction is quantized 
\begin{equation}
    p^{10}_j = \frac{\omega_j}{\sqrt{2}}(1-v_j^2) = \frac{N_j}{R_c} \hspace{30pt} \text{and} \hspace{30pt} q^{10}_s = \frac{\omega_s}{\sqrt{2}}(1-v_s^2) = \frac{N_s}{R_c}.
    \label{eqn: spatial quantization}
\end{equation}
While the above expression is universal in theories of gravity in flat space, it is a priori unclear whether the leading soft term $\mathcal{S}^{(-1)}$ is corrected upon compactification. For example, the leading soft term may be spoiled by $\mathcal{O}(1/R_c)$ corrections. In \cite{Marotta:2019cip}, it was argued that such terms do not arise when the soft particle is a KK zero-mode, with momentum $q^{10}_s = 0.$ One may also demonstrate that one need not worry about any such corrections, more generally.\footnote{One can see that compactifying on a circle of radius $R_c$ does not spoil the leading soft term with a Feynman diagrammatic argument. Feynman diagrams for QFTs on tori have been well-studied in \cite{Khanna:2011gf, Khanna:2014qqa}. The main difference is that $p^{10}$ is now quantized, so integrals over $p^{10}$ get replaced by discrete sums 
\begin{equation}
    \int \frac{dp^{10}}{2\pi} \mapsto \frac{1}{R_c} \sum_{n \in \mathbb{Z}} ,
\end{equation} where $p^{10} = 2 \pi n/R_c$ (here $p^\mu$ is a general momentum) The Feynman propagators in momentum space retain an identical form up to this substitution. Vertex interactions in the theory are also unchanged as these are features of a local QFT and do not depend on the global topology of the space. One may now reproduce the Feynman-diagrammatic derivation for the leading soft theorem in compactified space with these modifications (see \cite{Weinberg:1965nx} for the uncompactified case). Because the structure of the Feynman diagrams is essentially unmodified, the derivation for the leading soft theorem proceeds as usual, and we find that $\mathcal{S}^{(-1)}$ is also unmodified.} Thus, the expression \eqref{eqn: leading soft compactification} holds for theories of gravity even when they have a compactified dimension.

It will be convenient to rewrite the leading soft term, $\mathcal{S}^{(-1)}$ in terms of the quantum numbers $(\omega_j,v_j^I)$, 
\begin{equation}
    \mathcal{S}^{(-1)} = -2 \kappa \sum_{j=1}^n \eta_s \eta_j \frac{\omega_j}{\omega_s} \frac{e_{IJ}(v_s-v_j)^I (v_s-v_j)^J}{(v_s-v_j)^2}.
    \label{eqn: soft KK mode2}
\end{equation}At this point, we should still view the scattering amplitude as occurring in an 11-dimensional theory of gravity with one compactified dimension. The only constraint on the parameters $(\omega_j,v_j^I)$ is the quantization condition \eqref{eqn: spatial quantization}.

\subsubsection*{Soft Theorems for D0-Branes in Type IIA String Theory}

Because the leading soft term retains its universal form even when one of the directions is compactified, it follows that soft theorems may be applied to KK modes as well. Specifically, we will consider a situation where the soft particle has non-zero KK momentum, $q^{10}_s = N_s/R_c.$ The soft limit now reads
\begin{equation}
    \frac{\omega_s}{\omega_j} = \frac{N_s}{N_j} \frac{1-v_j^2}{1-v_s^2} \rightarrow 0. 
    \label{eqn: soft KK modes}
\end{equation}
We see that the limit $N_s/N_j \rightarrow 0$ (keeping the parameters $v_s,v_j$ fixed) recovers the soft limit for KK modes. 

Inserting Equation \eqref{eqn: soft KK modes} into the soft theorem \eqref{eqn: soft KK mode2}, we find the following soft theorem associated to scattering of graviton KK modes in the 10-dimensional type IIA string theory. Now, the parameter $N_j$ labels the KK mode of the corresponding particle,
\begin{equation}
    \mathcal{S}^{(-1)} = -2 \kappa \sum_{j=1}^n \eta_s \eta_j \frac{N_s}{N_j} \frac{1-v_j^2}{1-v_s^2} \frac{e_{IJ}(v_s-v_j)^I (v_s-v_j)^J}{(v_s-v_j)^2}.
    \label{eqn: soft KK mode3}
\end{equation}
The KK modes are just D0-brane bound states where the integer $N_j$ labels the number of D0-branes in a particular bound state. Thus, we have recovered a soft theorem for D0-branes where the soft limit is precisely $N_s/N_j \rightarrow 0$, just as it was in the BFSS case.

\subsubsection*{Soft Theorem for Non-Relativistic D0-Branes: The BFSS Limit}

Up to this point, we have been studying D0-brane scattering in IIA where the D0-branes may be relativistic. The BFSS matrix model concerns only non-relativistic D0-brane dynamics, however. Fortunately, it is known how to rescale the parameters in type IIA string theory to recover the BFSS action for these non-relativistic D0-branes \cite{Seiberg_1997}. Indeed, BFSS can be viewed as a particular scaling limit of weakly coupled type IIA string theory, and the scattering amplitudes of D0-brane bound states in the two descriptions should agree by definition -- this agreement is at the heart of our argument.\footnote{Concretely, the BFSS action is just a limit of the nonabelian DBI action describing $N$ D0-branes in type IIA string theory from which the D0-brane scattering amplitudes that we are discussing can ostensibly be computed.} 

For the leading soft term, $\mathcal{S}^{(-1)}$, of Equation \eqref{eqn: soft KK mode3} this rescaling amounts to taking the non-relativistic limit $v_j \mapsto \varepsilon v_j$ where $\varepsilon \rightarrow 0.$ In this non-relativistic limit, Equation \eqref{eqn: soft KK mode3} reads
\begin{equation}
    \mathcal{S}_{\text{BFSS}}^{(-1)} = -2 \kappa \sum_{j=1}^n \eta_s \eta_j \frac{N_s}{N_j} \frac{e_{IJ}(v_s-v_j)^I (v_s-v_j)^J}{(v_s-v_j)^2}.
    \label{eqn: soft KK mode4}
\end{equation}
This is precisely the form for the leading soft theorem that we previously conjectured for D0-brane scattering in BFSS via the duality dictionary! A similar computation may be considered to show equality at subleading order, $\mathcal{S}^{(0)}.$ Though this calculation did not involve working with the $N \times N$ matrix degrees of freedom familiar to BFSS, it is still a bona fide non-relativistic D0-brane calculation in type IIA string theory and, thus, gives strong evidence that the BFSS matrix model enjoys the soft theorems discussed in Section \ref{BFSSSGT}. Nevertheless, a direct computation using the D0-brane worldline Hamiltonian \eqref{eqn: BFSS Hamiltonian} would be illuminating.

Note that while this argument also relies on the equivalence between M-theory compactified on a small circle and type IIA string theory, it is qualitatively different from Seiberg's classic argument that the DLCQ of M-theory should be described by BFSS \cite{Seiberg_1997}. The latter invokes an infinitely large Lorentz boost which relates DLCQ M-theory to $\widehat{\text{M}}$-theory (see Section \ref{sec:backgroundfield}) compactified on a small spatial circle (and so their amplitudes, being Lorentz invariant, might plausibly agree). This is not the spirit of our argument -- we instead give a first-principle computation of the relevant amplitudes in spatially compactified $\widehat{\text{M}}$-theory without discussing amplitudes in DLCQ M-theory at all. This requires us to take great care in showing that any possible $\mathcal{O}(1/R_c)$ corrections vanish for the particular case of soft theorems but allows us to avoid discussing any thorny details about possible convergence issues when infinitely large boosts are involved. Once we know the relationship between D0-brane amplitudes with a soft D0-brane and those without one in IIA, taking the BFSS limit wherin the external D0-branes are non-relativistic recovers BFSS amplitudes tautologically.

\section{Derivation of Conservation Laws in the Matrix Model}
\label{appendix: conservation derivation}

In this appendix, we will detail how the leading and subleading soft theorem in the matrix model implies the conservation of infinitely many charges $\mathcal{Q}[f]$ and $\mathcal{Q}[Y^I]$ given in Sections \ref{sec: leading charges} and \ref{sec: subleading charges}. 

\subsubsection*{Shadow Transform Details}

The heart of this argument relies on an integral transformation known as the shadow transform. We define the shadow transform on the D0-brane creation operator $\mathcal{G}_{IJ}^{\text{in} \dagger}(N_s,v_s)$ as
\begin{equation}
    \widetilde{\mathcal{G}}^{\hspace{1pt} \text{in},\dagger}_{\Delta,IJ}(N_s,v_s) = \mathcal{N}_{\Delta} ~ \delta^{\{K}_{\{I} \delta^{L\}}_{J\}} \int d^9 w_s ~ \frac{\mathcal{I}_{K M}(v_s-w_s) \mathcal{I}_{L N}(v_s-w_s)}{(v_s-w_s)^{2(9-\Delta)}} ~ \mathcal{G}_{MN}^{\hspace{1pt} \text{in} \dagger}(N_s,u_s),
    \label{eqn: shadow transform}
\end{equation}
where $\{\cdot,\cdot\}$ denotes the symmetric traceless projection on the given indices, $\mathcal{I}_{IJ}(v) = \delta_{IJ} - 2 v_Iv_J/v^{\hspace{1.5pt}2}$ is the inversion tensor familiar from conformal field theory, and $\mathcal{N}_\Delta$ is a normalization coefficient which regulates the integral
\begin{equation}
    \mathcal{N}_\Delta = \frac{(\Delta + 1)(10-\Delta)\Gamma(\Delta)\Gamma(9-\Delta)}{\pi^9(\Delta-1)(8-\Delta)\Gamma(\tfrac{9}{2}-\Delta)\Gamma(\Delta-\tfrac{9}{2})}.
\end{equation}
Notice that applying the shadow transform to the D0-brane creation operator $\mathcal{G}_{IJ}^{\text{in} \dagger}(N_s,v_s)$ does not affect the number of D0-branes added to the state -- it will always be $N_s$. Rather, it simply smears the velocity of the D0-branes which are added against some kernel.
This shadow transform is especially useful in celestial conformal field theory and plays a large role in demonstrating that soft theorems are equivalent to conservation laws in general gauge and gravitational theories.

There will be two particular shadow transforms that we need to extract details about the leading and subleading soft theorems and to derive their associated conservation laws. In particular, we define the \textit{leading shadow transformed creation operator} and \textit{subleading shadow transformed creation operator} respectively as
\begin{equation}
    \begin{split}
        &\textbf{leading:} \hspace{42pt} \widetilde{\mathcal{G}}^{\hspace{1pt} (-1) \hspace{1pt} \text{in} \dagger}_{IJ}(N_s,v_s) = \lim_{\Delta \rightarrow \hspace{1pt} 1} ~\widetilde{\mathcal{G}}^{\hspace{1pt} \text{in} \dagger}_{\Delta,IJ}(N_s,v_s) \\
        &\textbf{subleading:} \hspace{30pt} \widetilde{\mathcal{G}}^{\hspace{1pt} (0) \hspace{1pt} \text{in} \dagger}_{IJ}(N_s,v_s) = \lim_{\Delta \rightarrow \hspace{1pt} 0} ~\widetilde{\mathcal{G}}^{\hspace{1pt} \text{in} \dagger}_{\Delta,IJ}(N_s,v_s).
    \end{split} 
\end{equation}
Shadow transforms for the outgoing creation operators may be defined in an identical way.

\subsection*{Derivation of Leading-Order Conserved Charges}

We begin by restating the soft theorem in the matrix model (Equation \eqref{eqn: leading BFSS})
\begin{equation}
    \begin{split}
        \frac{N_s}{R} \langle \text{out} | \frac{1}{2} \Big[ \mathcal{G}^{\hspace{1pt}(-1)\hspace{1pt} \text{out}}_{IJ}(&N_s,v_s) - \mathcal{G}^{\hspace{1pt}(-1) \hspace{1pt} \text{in} \hspace{1pt} \dagger}_{IJ}(N_s,v_s)\Big]|\text{in}\rangle \\
        &= \frac{\kappa}{2}\bigg[-2 \sum_{j=1}^n \eta_j \frac{N_j}{R} \frac{(v_s-v_j)^I(v_s-v_j)^J}{(v_s-v_j)^2}\bigg]\langle \text{out}|\text{in}\rangle + \mathcal{O}\Big(\frac{N_s}{R}\Big) \\
        &= \frac{\kappa}{2}\bigg[-2 \sum_{j=1}^n \eta_j \frac{N_j}{R} \frac{(v_s-v_j)^I(v_s-v_j)^J}{(v_s-v_j)^2}\bigg]\langle \text{out}|\text{in}\rangle + \mathcal{O}\Big(\frac{1}{N}\Big),
    \end{split}
\end{equation}
where we have used the fact that $R \sim N$ and $N_s \sim \mathcal{O}(1)$ to write $\mathcal{O}(N_s/R)$ and $\mathcal{O}(1/N)$ on the last line. It has been shown one may apply the shadow transform to both sides of this expression to find \cite{Kapec_2022}
\begin{equation}
    \begin{split}
        \frac{N_s}{4R} \langle \text{out} | \frac{1}{2} \partial_{v_s^I}\partial_{v_s^J} \Big[ \widetilde{\mathcal{G}}^{\hspace{1pt}(-1) \hspace{1pt} \text{out}}_{IJ}(&N_s,v_s) - \widetilde{\mathcal{G}}^{\hspace{1pt}(-1) \hspace{1pt} \text{in} \hspace{1pt} \dagger}_{IJ}(N_s,v_s)\Big]|\text{in}\rangle \\
        &= \frac{\kappa}{2}\sum_{j=1}^n \eta_j \frac{N_j}{R}\delta^{(9)}(v_s-v_j)\langle \text{out}|\text{in}\rangle + \mathcal{O}\Big(\frac{1}{N}\Big).
    \end{split}
\end{equation}
Though the right hand side of this expression has singular behavior when $v_s = v_j$, we may smooth out this feature by integrating against an arbitrary function $f(v_s),$
\begin{equation}
    \begin{split}
        \frac{N_s}{8R} \langle \text{out} |\int d^9 v_s ~ f(v_s) \partial_{v_s^I}\partial_{v_s^J} \Big[ \widetilde{\mathcal{G}}^{\hspace{1pt}(-1) \hspace{1pt} \text{out}}_{IJ}(&N_s,v_s) - \widetilde{\mathcal{G}}^{\hspace{1pt}(-1) \hspace{1pt} \text{in} \hspace{1pt} \dagger}_{IJ}(N_s,v_s)\Big]|\text{in}\rangle \\
        &= \frac{\kappa}{2}\sum \eta_j \frac{N_j}{R}f(v_j)\langle \text{out}|\text{in}\rangle + \mathcal{O}\Big(\frac{1}{N}\Big).
    \end{split}
\end{equation}
Moving the first term on the right-hand side to the left hand side yields an expression that vanishes as $\mathcal{O}(1/N)$. An immediate consequence is that the charges $\mathcal{Q}^{\text{in}}[f]$ and $\mathcal{Q}^{\text{out}}[f]$ defined in Equation \eqref{eqn: bfss supertranslation charge definition} satisfy the conservation law
\begin{equation}
    \langle \text{out}| \Big( \mathcal{Q}^{\text{out}}[f] - \mathcal{Q}^{\text{in}}[f] \Big) |\text{in}\rangle = \mathcal{O}\Big(\frac{1}{N}\Big).
\end{equation}

\subsubsection*{Derivation of Subleading-Order Conserved Charges}

The first set of conserved charges, $\mathcal{Q}[f]$, were derived by considering the leading term in the soft expansion only. To derive a second set, $\mathcal{Q}[Y^I]$, we wish to isolate the subleading term in the soft expansion. One could do this by applying the following projection operator $(1 + \tfrac{N_s}{R} \partial_{N_s/R})$ to the soft expansion -- this would project out the leading soft term but preserve the subleading corrections. However, $N_s \in \mathbb{N}$ is quantized, so taking derivatives is ill-defined. Instead, we will approximate this derivative operator by finite differences. Upon doing so, the projection operator reads
\begin{equation}
    \mathbb{P} \hspace{1pt} \mathcal{G}_{IJ}^{\hspace{1pt} \text{in} \hspace{1pt} \dagger}(N_s,v_s) = \mathcal{G}_{IJ}^{\hspace{1pt} \text{in} \hspace{1pt} \dagger}(N_s,v_s) + N_s\Big[\mathcal{G}_{IJ}^{\hspace{1pt} \text{in} \hspace{1pt} \dagger}(N_s + 1,v_s) - \mathcal{G}_{IJ}^{\hspace{1pt} \text{in} \hspace{1pt} \dagger}(N_s,v_s)\Big].
    \label{eqn: projection op}
\end{equation}

If we write the soft theorem schematically as
\begin{equation}
    \langle \text{out} | \frac{1}{2}\Big[ \mathcal{G}_{IJ}^{\hspace{1pt} \text{out}}(N_s,v_s) - \mathcal{G}_{IJ}^{\hspace{1pt} \text{in} \hspace{1pt} \dagger}(N_s,v_s)\Big] \text{in}\rangle = \frac{N_j}{N_s} 
    \mathcal{S}^{(-1)}\langle \text{out}|\text{in}\rangle + \mathcal{S}^{(0)}\langle \text{out}|\text{in}\rangle + \mathcal{O}\Big(\frac{1}{N}\Big),
\end{equation}
one may quickly verify the following relation among the projected soft D0-brane creation operators
\begin{equation}
    \begin{split}
        \langle \text{out} | \frac{1}{2}\Big[ \mathbb{P} \hspace{1pt} \mathcal{G}_{IJ}^{\hspace{1pt} \text{out}}(&N_s,v_s) - \mathbb{P} \hspace{1pt} \mathcal{G}_{IJ}^{\hspace{1pt} \text{in} \hspace{1pt} \dagger}(N_s,v_s)\Big] \text{in}\rangle \\
        &= \mathcal{S}^{(-1)} \bigg(\frac{N_j}{N_s^2}\frac{1}{1+1/N_s}\bigg) \langle \text{out}|\text{in}\rangle + \mathcal{S}^{(0)} \langle \text{out} |\text{in}\rangle + \mathcal{O}\Big(\frac{1}{N}\Big) \\
        &= \mathcal{S}^{(0)} \langle \text{out} |\text{in}\rangle + \mathcal{O}\Big(\frac{1}{N}\Big).
    \end{split}
    \label{eqn: projection derivation}
\end{equation}
Thus, we have projected out the leading piece, as desired. There is a slight technical complication, however, in that we must take $N_s \sim N^{\alpha}$ for some $\alpha \in (\tfrac{1}{2},1)$ so that the term $N_j/N_s^2(1+ 1/N_s)$ vanishes in the large $N$, M-theory limit. While this term won't quite vanish as fast as $1/N$, it does vanish in the M-theory limit, so we group it into the $\mathcal{O}(1/N)$ corrections.

    It has been shown that one may apply the shadow transform to both sides of this expression to find \cite{Kapec_2018}.
\begin{equation}
    \begin{split}
        \langle \text{out} | \frac{1}{2}\Big[&\partial_{v_s^I} \hspace{1pt} \mathbb{P} \hspace{2pt} \widetilde{\mathcal{G}}^{\hspace{1pt}(0) \hspace{1pt} \text{out}}_{IJ}(N_s,v_s) - \partial_{v_s^I} \hspace{1pt} \mathbb{P} \hspace{2pt} \widetilde{\mathcal{G}}^{\hspace{1pt}(0) \hspace{1pt} \text{in} \hspace{1pt} \dagger}_{IJ}(N_s,v_s)\Big]|\text{in}\rangle \\
        &= \frac{\kappa}{2}\sum_{j=1}^n \eta_j \bigg[\delta^{(9)}(v_s-v_j) \partial_{v_j^J} + \frac{i}{2} \partial_{v_s^I} \delta^{(9)}(v_s-v_j) S_j^{IJ} + \frac{1}{9} \partial_{v_s^J} \delta^{(9)}(v_s-v_j) \frac{N_j}{R} \partial_{N_j/R}\bigg] \\
        &\hspace{70pt} \times \langle \text{out}|\text{in}\rangle + \mathcal{O}\Big(\frac{1}{N}\Big),
    \end{split}
\end{equation}
where the projection operator, $\mathbb{P}$, acts on the shadow transformed creation operator as in Equation \eqref{eqn: projection op}. Now, we smear both sides of this expression against the arbitrary vector field, $Y^I$ to smooth over the delta-function singularity. The result is that the charges $\mathcal{Q}^{\text{in}}[Y^I]$ and $\mathcal{Q}^{\text{out}}[Y^I]$ defined in Equation \eqref{eqn: bfss superrotation charge definition} satisfy the conservation law
\begin{equation}
    \langle \text{out}| \Big( \mathcal{Q}^{\text{out}}[Y^I] - \mathcal{Q}^{\text{in}}[Y^I] \Big) |\text{in}\rangle = \mathcal{O}\Big(\frac{1}{N}\Big),
\end{equation}
where now the $\mathcal{O}(1/N)$ terms on the right hand side may vanish slightly slower than $1/N$, but still do vanish in the M-theory limit.

\section{Conservation Laws for 11d Poincar\'e Symmetry in M-Theory and BFSS}
\label{appendix: Poincare symmetry}

In this appendix, we will consider the conservation equations associated to the leading and subleading soft theorems in M-theory (Equation \eqref{eqn: M-theory conservation}) and BFSS (Equation \eqref{eqn: BFSS conservation})
\begin{equation}
    \begin{split}
        \textbf{M-theory:}& \hspace{40pt} \langle \text{out}| \Big(Q^{\text{out}}[f,Y^I] - Q^{\text{in}}[f,Y^I]\Big)|\text{in}\rangle = 0 \\
        \textbf{BFSS:}& \hspace{40pt} \langle \text{out}| \Big(\mathcal{Q}^{\text{out}}[f,Y^I] - \mathcal{Q}^{\text{in}}[f,Y^I]\Big)|\text{in}\rangle = \mathcal{O}\Big(\frac{1}{N}\Big).
    \end{split}
\end{equation}
As discussed in Section \ref{BFSSconservlaw}, for several choices of functions $(f,Y^I)$ the conservation laws reduce to simple expressions. In the remainder of this appendix, we provide these expressions and give their interpretations. These identities demonstrate that D0-brane scattering amplitudes in BFSS enjoy the full 11d Poincar\'e symmetry in the large $N$ limit.
\vspace{5pt}

\noindent For $f(v)=1$, $Y^K(v)\partial_K=0$,
\begin{equation}
    \begin{split}
        & \hspace{52pt} \mathlarger{\sum}_{j} \hspace{3pt} \eta_j \omega_j \mathcal{A}_{\text{M}} = 0 \hspace{165pt} (\text{conservation of $P^+$}) \\
        & \hspace{52pt}  \mathlarger{\sum}_{j} \hspace{3pt} \eta_j  \mathlarger{\frac{N_j}{R}} \mathcal{A}_{\text{BFSS}} = \mathcal{O}\Big(\frac{1}{N}\Big) \hspace{118pt} (\text{conservation of D0-branes})
        \label{eqn: poincare first}
    \end{split}
\end{equation}
For $f(v)=\sqrt{2} v^I$, $Y^K(v)\partial_K=0$,
\begin{equation}
    \begin{split}
        &\hspace{57pt} \mathlarger{\sum}_{j} \hspace{3pt} \eta_j \sqrt{2}\omega_j v^I_j \mathcal{A}_{\text{M}} = 0 \hspace{140pt} (\text{conservation of $P^I$}) \\
        &\hspace{57pt} \mathlarger{\sum}_{j} \hspace{3pt} \eta_j \sqrt{2}\mathlarger{\frac{N_j}{R}} v^I_j \mathcal{A}_{\text{BFSS}} = \mathcal{O}\Big(\frac{1}{N}\Big) \hspace{92pt} (\text{conservation of momentum})
    \end{split}
\end{equation}
For $f(v)=v^{\hspace{1pt}2}$, $Y^K(v)\partial_K=0$,
\begin{equation}
    \begin{split}
        & \hspace{35pt} \mathlarger{\sum}_{j} \hspace{3pt} \eta_j \omega_j v^2 \mathcal{A}_{\text{M}} = 0 \hspace{155pt} (\text{conservation of $P^-$}) \\
        & \hspace{35pt} \mathlarger{\sum}_{j} \hspace{3pt} \eta_j \mathlarger{\frac{N_j}{R}} v^2 \mathcal{A}_{\text{BFSS}} = \mathcal{O}\Big(\frac{1}{N}\Big) \hspace{108pt} (\text{conservation of energy)}
    \end{split}
\end{equation}
For $f(v)=0$, $Y^K(v)\partial_K = -(v^I \partial_J - v^J \partial_I)$,
\begin{equation}
    \begin{split}
        & \mathlarger{\sum}_{j} \hspace{3pt} \eta_j [-i(v_j^I\partial_{v_j^J}-v_j^J\partial_{v_j^I})+ S_j^{IJ}]\mathcal{A}_{\text{M}} = 0 \hspace{52pt} (\text{conservation of $J_{IJ}$})\\
        & \mathlarger{\sum}_{j} \hspace{3pt} \eta_j [-i(v_j^I\partial_{v_j^J}-v_j^J\partial_{v_j^I})+ S_j^{IJ}]\mathcal{A}_{\text{BFSS}} = \mathcal{O}\Big(\frac{1}{N}\Big) \hspace{10pt} (\text{$SO(9)$ Invariance})
    \end{split}
\end{equation}
For $f(v)=0$, $Y^K(v)\partial_K = \frac{1}{\sqrt{2}}\partial_I$,
\begin{equation}
    \begin{split}
        & \hspace{60pt} \frac{1}{\sqrt{2}}\mathlarger{\sum}_{j} \hspace{3pt} \eta_j i\partial_{v_j^I} \mathcal{A}_{\text{M}} = 0 \hspace{138pt} (\text{conservation of $J_{-I}$})\\
        & \hspace{59pt} \frac{1}{\sqrt{2}}\mathlarger{\sum}_{j} \hspace{3pt} \eta_j i\partial_{v_j^I} \mathcal{A}_{\text{BFSS}} = \mathcal{O}\Big(\frac{1}{N}\Big) 
        \hspace{97pt} (\text{Galilean boost invariance})
    \end{split}
\end{equation}
For $f(v)=0$, $Y^K(v)\partial_K = v^K\partial_K$,
\begin{equation}
    \begin{split}
        & \hspace{40pt} \mathlarger{\sum}_{j} \hspace{3pt} \eta_j  i(v_j^I\partial_{v_j^I} - \omega_j \partial_{\omega_j}) \mathcal{A}_{\text{M}} = 0 \hspace{90pt} (\text{conservation of $J_{+-}$})\\
        & \hspace{40pt} \mathlarger{\sum}_{j} \hspace{3pt} \eta_j  i(v_j^I\partial_{v_j^I} - \mathlarger{\frac{N_j}{R}} \partial_{N_j/R}) \mathcal{A}_{\text{BFSS}} = \mathcal{O}\Big(\frac{1}{N}\Big) \hspace{32pt} (\text{new symmetry})
    \end{split}
\end{equation}
For $f(v)=0$, $Y^K(v)\partial_K = \frac{1}{\sqrt{2}}(v^2 \partial_I-2v^I v^K \partial_K)$,
\begin{equation}
    \begin{split}
        & \frac{1}{\sqrt{2}} \mathlarger{\sum}_{j} \hspace{3pt} \eta_j  [i (v_j^2 \partial_{v_j^I}-2v_j^I v_j^J \partial_{v_j^J} + 2v_j^I \omega_j \partial_{\omega_j} ) + 2 v_j^J S_j^{IJ} ] \mathcal{A}_{\text{M}} = 0 \hspace{46pt} (\text{conservation of $J_{+I}$})\\
        & \frac{1}{\sqrt{2}} \mathlarger{\sum}_{j} \hspace{3pt} \eta_j  [i (v_j^2 \partial_{v_j^I}-2v_j^I v_j^J \partial_{v_j^J}  ) + 2v_j^I \mathlarger{\frac{N_j}{R}} \partial_{N_j/R} + 2 v_j^J S_j^{IJ} ] \mathcal{A}_{\text{BFSS}} = \mathcal{O}\Big(\frac{1}{N}\Big) \hspace{12pt} (\text{new symmetry})
        \label{eqn: poincare - last}
    \end{split}
\end{equation}

Note that the BFSS conservation laws follow from Equation \eqref{eqn: BFSS conservation} which is a consequence of the soft theorems discussed in Section \ref{sec: soft theorems} and derived in Appendix \ref{appendix: first principles argument}. It follows that these symmetries are actually manifested in the matrix model; our derivation does not presuppose the BFSS duality relation among scattering amplitudes. Note the striking similarity between the top and bottom lines of Equations \eqref{eqn: poincare first}-\eqref{eqn: poincare - last}. The BFSS conservation laws are nothing but the M-theory conservation laws after translating through the duality dictionary. This gives definitive evidence that BFSS scattering amplitudes enjoy 11$d$ Poincar\'e symmetry.

\bibliography{mybib.bib}
\bibliographystyle{apsrev4-1long}

\end{document}